\documentclass[12pt]{article}

\usepackage{amssymb}
\usepackage{amsfonts}
\usepackage{mathrsfs}
\usepackage{latexsym}
\usepackage{amsmath}
\usepackage{graphics}
\usepackage{graphicx}

\chardef\tempcat=\the\catcode`\@ \catcode`\@=11
\def\cyracc{\def\u##1{\if \i##1\accent"24 i%
    \else \accent"24 ##1\fi }}
\newfam\cyrfam
\font\tencyr=wncyr10
\def\cyr{\fam\cyrfam\tencyr\cyracc}

\newcommand{\ex}{\mathop{\mathrm{e}}\nolimits}
\newcommand{\Tr}{\mathop{\mathrm{Tr}}\nolimits}
\newcommand{\Ref}[1]{(\ref{#1})}

\newcommand{\mbf}[1]{\mbox{\boldmath{$#1$}}}
\newcommand{\dd}{\mathrm{d}}

\newcommand{\alg}{\mathfrak}
\newcommand{\diag}{\mathop{\mathrm{diag}}\nolimits}
\newcommand{\pd}{\partial}
\newcommand{\Prep}{\mathcal{F}}
\newcommand{\I}{\mathbb{I}}
\newcommand{\bJ}{\mathbb{J}}
\newcommand{\A}{\mathcal{A}}
\newcommand{\B}{\mathcal{B}}
\newcommand{\D}{\mathcal{D}}
\newcommand{\R}{\mathcal{R}}
\newcommand{\Real}{\mathbb{R}}
\newcommand{\Compl}{\mathbb{C}}

\newcommand{\mr}{\mathrm}
\newcommand{\T}{\mr{T}}
\newcommand{\eps}{\varepsilon}
\newcommand{\Pol}{\mathcal{P}}
\newcommand{\vpint}{\displaystyle -\hskip -1.1em \int}
\renewcommand{\k}{\mbf{k}}
\newcommand{\N}{\mathcal{N}}
\renewcommand{\Im}{\mathop{\mr{Im}}\nolimits}

\def\np#1#2#3{Nucl. Phys. {\bf B#1} (#2) #3}

\def\cmp#1#2#3{Comm. Math. Phys. {\bf #1} (#2) #3}

\def\bX{{\mathbb X}}
\def\bS{{\mathbb S}}

\def\bP{{\mathbb P}}
\def\bR{{\mathbb R}}
\def\bC{{\mathbb C}}
\def\bZ{{\mathbb Z}}

\def\bI{{\mathbb I}}
\def\bT{{\mathbb T}}

\def\zb{{\bar z}}

\def\CF{{\mathcal F}}

\def\CM{{\mathfrak M}}
\def\CN{{\mathcal N}}
\def\CO{{\mathcal O}}
\def\CP{{\mathcal P}}

\def\CV{{\mathcal V}}
\def\CW{{\mathcal W}}

\def\p{\partial}

\def\dd{{\rm d}}

\def\zb{\bar{z}}

\def\Tr{{\rm Tr}}

\def\dim{{\rm dim}}

\def\a{{\alpha}}

\def\b{{\beta}}
\def\d{{\delta}}

\def\ve{{\varepsilon}}
\def\vf{{\varphi}}
\def\m{{\mu}}

\def\u{{\Upsilon}}

\begin{document}
\makeatletter
\begin{titlepage}
\begin{flushright}
{hep-th/0404225}\\
{IHES/P/04/19}\\
{ITEP-TH-17/04}
\end{flushright}
\begin{center}
\vskip .5in
{\LARGE\bf ABCD of instantons}\\
\vskip 2cm

{\bf Nikita Nekrasov\footnote{on leave of absence from ITEP, Moscow, Russia} and Sergey Shadchin}
\end{center}
\bigskip
\bigskip
\centerline{\em Institut des Hautes Etudes Scientifiques}
\centerline{\em 35 route de Chartres, 91440 Bures-sur-Yvette, FRANCE}
\centerline{\tt email: nikita@ihes.fr, chadtchi@ihes.fr}
\abstract{\noindent We solve ${\CN}=2$ supersymmetric Yang-Mills theories
for
arbitrary classical gauge group, i.e. $SU(N)$, $SO(N)$, $Sp(N)$.
In particular, we derive the prepotential of the low-energy
effective theory, and the corresponding Seiberg-Witten curves.
We manage to do this without
resolving singularities of the compactified instanton moduli spaces.}
\bigskip
\end{titlepage}
\makeatother

\tableofcontents
\vfill\eject

\section{Introduction}

${\CN}=2$ supersymmetric Yang-Mills theory provides a unique laboratory for
studying Wilsonian effective action. The perturbation theory is rather
simple, and non-perturbative effects are crucial in order for the effective
theory to be well-defined, yet they are tractable  \cite{ABC}.
This effective action is an expansion in derivatives and fermions. The
leading, two-derivative, and up to four fermions piece
can be expressed with the help of a holomorphic function $\Prep(a)$,
known as prepotential  \cite{Prepotential}.
This function contains information about both perturbative (one loop)
and nonperturbative (instanton) corrections to
Wilsonian effective action of the theory.
Some ten years ago, the bold proposal was put forward by
Seiberg and Witten  \cite{SeibergWitten}, in case of $SU(2)$ gauge group, and then generalized to other Lie groups
 \cite{LiterSW}. According to this proposal, which we shall refer to as
Seiberg-Witten theory,
the prepotential is determined indirectly with the help of the symplectic geometry of a family
of certain algebraic curves. More precisely, the central charges of BPS
representations of ${\CN}=2$ superalgebra are the flat sections of a local
system over the moduli space of vacua. This system is associated to the
family of algebraic curves.
One associates to each point of
the moduli space of vacua an auxiliary complex curve and a
meromorphic differential in such a way that both the argument and the first
derivative of the prepotential become the periods of the differential.
By
carefully studying the monodromies of the prepotential and using its analytic
properties, most importantly those which follow from electric-magnetic
duality of (super)-Maxwell theory in four dimensions one can guess the
required family of curves in many examples. However, given a theory, i.e.
the gauge group and the matter representation, it is not a priori clear that
this guess is unique, and always exists.

Another way to get the curve is by NS5-D4 brane engineering and the subsequent
 M-theory lift,
which implies that the curve is, in fact,
a subspace of the world volume of a M5-brane  \cite{PrepFromM}. Again, this
approach is limited, since only very special matter representations are
realizable in this context, and only classical gauge groups. Another,
approach, often related to that one, is to realize gauge theory
via so-called geometrical
engineering,
using closed string backgrounds
 \cite{GeoEng}. In this approach the prepotential can be calculated by
doing
worldsheet instanton sum.

Recently another, much more direct and straightforward, technique was proposed \cite{SWfromInst}.

In this approach the prepotential is expressed as a sum of integrals of a well-defined
form over
instantons moduli spaces. More precisely, the gauge theory is considered in
a weak ${\CN}=2$ supergravity background, and the prepotential plus an
infinite set of corrections, which vanish in the limit of flat background,
are equal to the logarithm of the sum of instanton integrals. These
integrals can be evaluated using various techniques. In particular, the
method
 of  \cite{DeformationInstanton} reduces them to a multiple contour integral.
In the case of the gauge group $SU(N)$ these integrals were evaluated by
taking the sums of residues. The poles turn out to be in one-to-one
correspondence with the fixed points of a certain global symmetry group
action on the resolved instanton moduli space (the moduli space of
noncommutative $U(N)$ instantons). This point was the stumbling
block for further progress for some time, since it is well-known that there
is no such resolution of the instanton moduli space for other gauge groups.

Aslo in the previous work the precise definition of the contour integrals was
never spelled out very clearly. In this paper we correct this.

We shall show that the simplest approach to calculations of the
prepotential (and higher order gravitational couplings) goes via
lifting the gauge theory to five dimensions, compactifying it on a circle,
and then (if need be) taking the radius to zero  \cite{5dNik}.
In this approach the partition function of the gauge theory
turns out to be the character of the action of global gauge
transformations and the group of rotations on the space of
{\it holomorphic} functions on the moduli space of instantons.
In this approach the issue of compactification of the moduli space of
instantons does not even really arise. Indeed, the questions whether to add
point-like instantons, and whether to resolve the singularities of
the resulting space  do not arise, since these objects sit in high
complex codimension and holomorphic functions do not feel them.

This method gives the prepotential itself.
The prepotential is obtained in two forms:
as a formal power series over the dynamically generated scale $\Lambda$,
where
analytic properties are not evident, yet computationally each instanton term is rather
easy to get; and in the form of Seiberg-Witten geometry, where everything is
packaged as we described above, and analytic continuation is clear.
We
are able to compare our results with other results, existing in the literature.
In the case of the $SU(N)$ gauge group
the program of extracting Seiberg-Witten data from the $\Lambda$-expansion
was completed in  \cite{SWandRP} by using the saddle point method to sum up
the instanton series.

In this paper we extend the method of  \cite{SWfromInst},
developed initially for $SU(N)$ gauge group,
to the case of other classical groups, namely $SO(N)$ and $Sp(N)$.
We also obtain the Seiberg-Witten description of the supersymmetric Yang-Mills theories for this groups, and, in particular, rederive the relation
between the prepotential for $SU(N)$ theory and the prepotential for $SO(N)$ and $Sp(N)$ theories
 \cite{SOandSp,LecturesSWandIntegr}.

The paper is organized as follows:
 in section \ref{Outline} we collect some relevant information about the $\mathcal{N} = 2$
supersymmetric Yang-Mills theory. We then explain our strategy of the
calculation of the prepotential, following  \cite{SWfromInst}, in essence.

In section \ref{ADHM} we describe in some details
the construction of the instanton moduli space (the ADHM construction  \cite{ADHM}),
especially for the cases of $SO(N)$ and $Sp(N)$ groups.

In section section \ref{Derivation} we derive the formal expression for the preotential from the five dimensional perspective and then pull it back to the four dimensional theory.

In section \ref{SW} we derive Seiberg-Witten geometry, and the prepotentials. And finally in section \ref{check} we perform a consistency check of our results and compare them against the known expressions.


\subsection{Notations and conventions}

We study gauge theory on ${\bR}^4$. Sometimes it is convenient to compactify
${\bR}^4$ by adding a point $\bf \infty$ at infinity, thus producing
${\bS}^4 = {\bR}^4 \cup {\bf \infty}$.

We consider a principal $G$-bundle over ${\bS}^4$, with $G$ being one of the
classical groups ($SU(N)$, $SO(N)$ or $Sp(N)$).
To make ourselves perfectly clear we stress that $Sp(N)$ means
in this paper the group of matrices $2N \times 2N$
preserving the symplectic structure, sometimes denoted in the literature as
$USp(2N)$.

Let $A = A_\mu \dd x^\mu$ be a connection on the bundle. The curvature of this connection is
$$F = \frac{1}{2}F_{\mu\nu}\dd x^\mu \wedge \dd x^\nu = \dd A + A \wedge A.$$

The following convention will be used:
\begin{itemize}
\item The greek indices $\mu, \nu, \dots$ run over $0,1,2,3$, the small latin indices $i,j,\dots$ run over $1,2,3$,  the capital latin indices $A, B, \dots$ run over $1,2$,
\item the matrix representation of the quaternion algebra is:
$$\sigma_{\mu,\alpha\dot{\alpha}} = (\I_2, - i \tau_1, -i \tau_2, -i \tau_3)$$
where $\tau_i$ are Pauli matrices,
\item $\I_n$ is $n \times n$ unit matrix, and
$$
{\bJ}_{2n} = \left(
\begin{array}{cc}
 0 & {\I}_n \\
 -{\I}_n  & 0
\end{array}
\right)
$$
\item in the Minkowski space two homomorphisms $SL(2,\Compl) \to SO(3,1)$ are governed by
$$
\begin{aligned}
\sigma^\mu_{\alpha\dot{\alpha}} =& (\I_2, - \tau_1, -\tau_2, -\tau_3), \\
\bar{\sigma}^\mu_{\alpha\dot{\alpha}} =& (\I_2, + \tau_1, +\tau_2,+ \tau_3).
\end{aligned}
$$
(we apologise for the confusing notations -- we can only hope that every time
it will be clear
 whether we work with Euclidean  or Minkowski signature),
\item the generators of the spinor representation of $SO(3,1)$ are
$$ \sigma^{\mu\nu} = \frac{1}{4} \Big( \sigma^\mu \bar{\sigma}^\nu - \sigma^\nu \bar{\sigma}^\mu \Big),$$
they satisfy
$$
\sigma^{\mu\nu,\alpha\beta}\sigma^{\rho\sigma}_{\alpha\beta} = \frac{1}{2}\Big( g^{\mu\rho}g^{\nu\sigma} - g^{\mu\sigma}g^{\nu\rho} \Big) - \frac{i}{2}\epsilon^{\mu\nu\rho\sigma}.
$$
In the Euclidean space $\sigma^{\mu\nu}$ becomes the 't Hooft anti-selfdual projector,
\item the spinor metric is
$$\epsilon = \Vert \epsilon_{\alpha\beta} \Vert = \left(
\begin{array}{cc}
0 & -1 \\
1 & 0
\end{array}
\right),
$$
\item in the Euclidean space the complex conjugation rises and lowers the spinor indices without changing their dottness. In the Minkowski space the height of
the index is unchanged whereas its dottness does change.
\item ${\ve} = {\ve}_1 + {\ve}_2$

\item Mostly we denote by $G$ the gauge group. Sometimes,
when it does not lead to confusions $G$ also denotes its complexification
$G_{\bC}$. ${\frak g} = Lie(G)$. ${\bT} \subset G$ -- maximal torus,
$\frak t = Lie ({\bT})$ -- Cartan subalgebra, $W$ -- Weyl group, $h^\vee$ -- the dual Coxeter number.
We use the notations ${\vec a}$ or simply $a$ for the elements of $\frak t$.

For Lie group $H$, ${\bT}_{H}$ denotes its Cartan subgroup, $W_H$ -- Weyl.

$\Delta$ denotes the set of
roots of $\frak g$, or $\frak h$,
$\Delta^+ \subset \Delta$ -- the set of positive
roots. For $\a \in \Delta$, $\phi \in \frak t$, $\langle \a , \phi \rangle$
denotes the value of the root on $\phi$. For $h = \ex^{i \phi} \in \bT$, $h^{\a} \in
U(1)$ denotes ${\exp} i \langle \a , \phi \rangle$.

$G^D_k$ denotes dual (in the sense of instanton
reciprocity) Lie group (see the definition in the end of the section \ref{ADHMSU}), ${\bT}_k \subset G_k^D$ its maximal torus,
$W_k$ its Weyl group.

\end{itemize}


\section{$\mathcal{N} = 2$ Super Yang-Mills}
\label{Outline}

In this section we give a short outline of known
facts about $\mathcal{N} = 2$ supersymmetric Yang-Mills theory, the twist which makes it a topological field theory,
and the prepotential.


\subsection{The field content and the action}

The fields of ${\CN}=2$ super-Yang-Mills theory fall into the representations of the superalgebra  \cite{SUSYandSUGRA}
with eight
supercharges. The field content of the minimal $\N = 2$ super-Yang-Mills theory is the following:
$A_\mu, \psi_\alpha^A$ and $\phi$, where $A_\mu$ is a vector boson,
$\psi_\alpha^A$, $A = 1,2$ are two Weyl spinors and $\phi$ is a complex scalar.
Since vector bosons are usually associated with a gauge symmetry,
$A_\mu$ is supposed to be a gauge boson corresponding to a gauge group $G$.
It follows that it transforms in the adjoint representation of $G$.
To maintain the $\N = 2$ supersymmetry $\psi_\alpha^A$ and $\phi$ should also transform in the
adjoint representation. These fields form the ($\N$ = 2) chiral multiplet (sometimes called the gauge or the vector multiplet).

The most natural
superfield representation for the chiral multiplet is given
 in the extended superspace, which has the coordinates $x^\mu, \theta^\alpha_A, \bar{\theta}_{\dot{\alpha}}^A$, $A = 1,2$.
Then we have
\begin{equation}
\label{SUFI}
\Psi(x,\theta,\bar{\theta}) = \phi(x) + \sqrt{2}\theta_A^\alpha\psi^A_\alpha(x) - \frac{i}{\sqrt{2}}\epsilon_{AB}\theta^{\alpha A}\theta^{\beta B}\sigma_{\alpha\beta}^{\mu\nu} F_{\mu\nu}(x) + \dots
\end{equation}

The expression for the microscopic action of the $\N = 2$ is written as follows:
\begin{equation}
\label{MicroAction}
S = \frac{1}{8\pi h^\vee} \int d^4x d^4\theta \Im \left( \frac{\tau}{2} \Tr_{\mr{adj}} \Psi^2 \right)
\end{equation}
where $\displaystyle\tau = \frac{4\pi i}{g^2} + \frac{\Theta}{2\pi}$, $g^2$ being the Yang-Mills coupling constant (and the Plank constant as well) and $\Theta$ is the instanton
angle. Its contribution to the action is given by the topological term, $\Theta k$ where $k$ is the
instanton number:
\begin{equation}
\label{InstNumb}
k = - \frac{1}{16\pi^2 h^\vee} \int \Tr_{\mr{adj}} (F \wedge F),
\end{equation}
where $\Tr_{\mr{adj}}$ means that the trace is taken over the adjoint representation.

In the low energy limit, when the $\N=2$ supersymmetry is unbroken, the most general effective action can be obtained by the following generalization of \Ref{MicroAction}:
$$
S_{\mr{eff}} = \frac{1}{8\pi h^\vee} \int d^4 x d^4 \theta \Im \Big[  \Prep(\Psi , \Lambda)\Big]
$$
where $\Prep(a,\Lambda)$ is a holomorphic gauge-invariant function called the prepotential.
Its classical expression can be read form $\Ref{MicroAction}$: $\displaystyle\Prep_{\mr{class}}(a) = \frac{\tau}{2} a^2$.
All perturbative correction are contained in the 1-loop term which is equal to
\begin{equation}
\label{pertPrep}
\Prep_{\mr{pert}}(a,\Lambda) = - \sum_{\alpha \in {\Delta}^+} {(\alpha \cdot a)}^2 \left( \ln \left| \frac{\alpha \cdot a}{\Lambda} \right| - \frac{3}{2}\right)
\end{equation}
where $\Lambda$ is
the dynamically generated scale. In this formula the highest root is supposed to have length 2.


\subsection{Seiberg-Witten theory}

The non-perturbative part of the prepotential can be written as an infinite series over $\Lambda$
\begin{equation}
\label{instPrep}
\Prep_{\mr{inst}}(a,\Lambda) = \sum_{k=1}^\infty \Prep_k (a) \Lambda^{2h^\vee k}.
\end{equation}
These terms give rise to the instanton corrections to Green's functions.
However the direct calculation of their contribution is very complicated,
thus making quite useful the Seiberg-Witten theory  \cite{SeibergWitten}.
Accounting for monodromies of the prepotential one can show that the
prepotential can be expressed using the following data:
the auxiliary algebraic curve  and a meromorphic differential $\lambda$ defined on
it, such that its variations are holomorphic differentials.

If the rank of the gauge group is $r$ one requires $2r$
nontrivial cycles on this curve $A_1,\dots,A_r$ and $B_1,\dots,B_r$ such that
\begin{equation}
\label{SWdata}
\begin{aligned}
a_l &= \frac{1}{2\pi i}\int_{A_l}\lambda  & \ \ &\mbox{and} & \ \ \frac{\pd \Prep}{\pd a_l}
&= \frac{1}{2\pi i}\int_{B_l} \lambda,
& \ l &= 1,\dots r.
\end{aligned}
\end{equation}
The more detailed explanation of the Seiberg-Witten theory can be found, for example, in  \cite{LecturesSWandIntegr}.


\subsection{Topological twist}

In  \cite{TQFT} it was shown that the $\N=2$ supersymmetric Yang-Mills theory can be reformulated in such a way that it becomes a topological filed theory.
Namely, the action \Ref{MicroAction}, up to a term, proportional to ${\Tr} F
\wedge F$,
can be rewritten as a $\bar{Q}$-exact expression for a
BRST-like operator $\bar{Q}$ satisfying the
BRST-like property: $\bar{Q}^2 = 0$.
One can construct this operator by twisting the usual
 supersymmetry generators $\bar{Q}_{\dot{\alpha} A}$ in the following way:
$$
\bar{Q} = \epsilon^{\dot{\alpha}A}\bar{Q}_{\dot{\alpha} A}.
$$

If one considers an observable which is $\bar{Q}$ closed, then after the Wick rotation the functional integral
representing the vacuum expectation of
the observable localizes on the moduli space of the solutions of anti-self-duality equation
$$
F_{\mu\nu} = - \frac{1}{2}\epsilon_{\mu\nu\rho\sigma}F_{\rho\sigma} = - \star F_{\mu\nu}
$$
e.g. on the instanton moduli space. (In the Euclidean space-time ${\bR}^4$,
where we land after the Wick rotation, we don't care about the upper and lower indices).

It proves useful to consider
the following deformation of the BRST-like operator. One defines  \cite{SWfromInst}
$$
\begin{aligned}
Q^\mu &= Q^{\alpha A}\sigma_{\alpha A}^\mu &\  \ &\mbox{and}\ \ &
\bar{Q}_\eps = \bar{Q} + \Omega_{\mu\nu} x^\mu Q^\nu
\end{aligned}
$$
where $\Omega_{\mu\nu}$ is the matrix of infinitesimal rotation of the
${\bR}^4$.
There is coordinate system on ${\bR}^4$, in which it has the canonical form:
\begin{equation}
{\Omega} = \left(
\begin{array}{c}
\ {\eps}_1 \cdot {\bJ}_2 \qquad \; \quad {\bf 0} \quad\\
\quad  {\bf 0} \quad \; \qquad {\eps}_2 \cdot {\bJ}_2 \\
\end{array}
\right),
\label{omg}
\end{equation}
The story above can be retold using twisted superspace as follows: we introduce the supercoordinates
$\bar{\theta} = \epsilon_{\dot{\alpha} A} \bar{\theta}^{\dot{\alpha} A}$
and $\theta^\mu$ according to
$\theta^{\alpha A} = \theta^\mu \sigma_\mu^{\alpha A}$.
Using this notation the superfield \Ref{SUFI} can be rewritten as follows:
$$
\Psi(x,\theta,\bar{\theta}) = \phi(x) + \sqrt{2}\theta^\mu \psi_\mu - \frac{i}{\sqrt{2}} \theta^\mu\theta^\nu \Big(F_{\mu\nu}  - i \star F_{\mu\nu} \Big) + \dots
$$
where $\psi_{\alpha A} = \psi_\mu \sigma^\mu_{\alpha A}$. It worth noting that the term in the brackets after the Wick rotation becomes $F_{\mu\nu} - \star F_{\mu\nu} = 2 (F_{\mu\nu})^-$, that is, the anti-selfdual part of the curvature.

The action \Ref{MicroAction} is not $\bar{Q}_\eps$ exact. However, one can deform this action in such a way, that new action will be $\bar{Q}_\eps$ exact. This action can be considered as a $5+1$ dimensional $\N=1$ supersymmetric Yang-Mills theory
in the so-called $\Omega$-background and
compactified on the two dimensional torus.


\section{The prepotential}

In  \cite{SWfromInst} it was shown
that the partition function of the ${\CN}=2$ gauge theory  in
$\Omega$-background is closely related to
the prepotential of effective theory.

The logic of the identification is the following. The partition function
is obtained by integrating out all degrees of freedom, except for the
zero mode of the Higgs field ${\Phi}$. We can consider doing this
integration
in two steps -- first, down to some low energy scale ${\mu}$ and then
all the way from ${\mu}$ to zero. At the energy scale ${\mu}$ we would have
an expression:
\begin{equation}
Z(a, {\Lambda}; {\ve}_1, {\ve}_2 ) = \int_{\rm low \ energy \ fields}
{\exp} \int {\rm d}^4 x \ {\rm d}^4 {\theta} \ {\CF}\left({\Psi}(x,  {\theta}),
{\Lambda}(x, {\theta})\right) + \ldots
\label{eff}
\end{equation}
where $\ldots$ represent higher derivative terms. Here ${\Lambda}(x,
{\theta})$ is an effective low-energy scale, which in the
$\Omega$-background becomes a function on the superspace. The
supercharge ${\bar Q}_{\ve}$, preserved by the $\Omega$-background act as derivatives
${\theta} {\p\over\p x} - {\Omega}x {\p\over\p\theta}$. We can add to the
action ${\bar Q}_{\ve}$-exact term, which would localize the integal over
the superspace to the vicinity of the point $x = 0, {\theta} = 0$ -- the origin.
Similarly, one can argue that the whole path integral would localize on the
fields, which are concentrated near the origin.
This line of arguments leads to the expansion:
\begin{equation}
Z (a, {\Lambda} ; {\ve}_{1}, {\ve}_2 ) = \exp \frac{1}{{\ve}_1{\ve}_2}
{\CF} ( a; {\Lambda} ) + \ldots
\label{Z}
\end{equation}
where $\ldots$ denote less singular terms. They are also
meaningful  \cite{SWfromInst}, but in this paper we shall touch upon them.

\subsection{Five dimensional perspective}

Things start looking much nicer and simpler
if we take a higher dimensional view.

Consider five dimensional gauge theory with {\bf 8} supercharges,
 the lift of the ${\CN}=2$ theory.
Compactify it on a circle, of circumference ${\beta}$,
${\bS}^1_{\b}$. Let us denote by $y$ the coordinate on the circle,
$ y \sim y   + {\b}$.
In addition, impose the twisted boundary conditions on the (non-compact)
spatial part. We work with Euclidean signature of the four dimensional
space. The twisted boundary conditions consist of identifications
$(y, {\vec x}) \sim ( y + {\b}, {\exp} {\b} {\Omega} \cdot {\vec x})$,
where ${\Omega}$ is the generator \Ref{omg} of infinitesimal $SO(4)$ rotation.
More on
the definition of the $\Omega$-background
and further explanation can be found in  \cite{SmallInst,SWandRP}.

In  \cite{5dNik} it is explained that in the limit of small bare gauge
coupling the pure ${\CN}=2$ five dimensional gauge theory with gauge group $G$, in the sector
with instanton charge $k$ reduces to the
supersymmetric quantum mechanics on the instanton moduli space ${\CM}_{G,k}$.
Now, having compactified five dimensional theory on the circle together with the
twist in the four dimensional part translates to the setup for the
calculation of twisted Witten index in
the supersymmetric quantum mechanics.
Minimal supersymmetric quantum  mechanics calculates index of Dirac
operator (for more on the relation between Dirac operator, supersymmetric
quantum mechanics, and their appearences in gauge theories,
see  \cite{sqm}). Having fixed some complex structure on ${\bR}^4 \approx {\bC}^2$ we
can translate this to the calculation of the index of
appropriately twisted
${\bar\p}$ operator on the moduli space of instantons on ${\bR}^4$:
$$
{\Tr}(-1)^F \sim \sum_{p} (-)^{p} {\Tr}_{H^{0,p}({\CM}_{G,k})} {\bf g}
$$
where ${\bf g}$ is the element of the global symmetry group,
which in the case of the problem at hands is the product of the group
of global gauge transformations, i.e. a copy of $G$ and the group of
Euclidean rotations, i.e. $SO(4)$. More precisely, if we are to use the complex
structure, we should reduce $SO(4)$ down to $U(2)$. However, as the trace
depends only on the conjugacy class, the difference between these two is
immaterial. So ${\bf g}  = ( h , {\exp}{\b}{\Omega} )$. We assume that the element $h$
of $G$ is generic. Let ${\bT}$ be the
corresponding maximal torus, containing it. Let ${\vec a} \in \frak{t} \equiv Lie({\bT})$
be such, that ${\exp}({\b}{\vec a}) = h$.

This reasoning leads to the formula:
\begin{equation}
Z_{5}({\vec a}, {\Lambda}_5 , {\ve}_1 , {\ve}_2 ; {\b}) Z_{5}^{pert}({\vec a}, {\Lambda}_5 , {\ve}_1, {\ve}_2; {\b}) \times
\sum_{k=0}^{\infty}
{\Lambda}^{k}_{5}\  {\rm S}{\Tr}_{H({\CM}_{G,k})} e^{{\b}{\vec a}\cdot {\vec T}}
e^{{\b}{\Omega}}
\label{part}
\end{equation}
Here ${\Lambda}_5$ simply counts the instanton number.
It is proportional to
${\Lambda}_{\rm uv}^{2h^{\vee}} \ex^{-\frac{8{\pi}^2 {\b}}{g_5^2}}$, where $g_5$ is
the bare five dimensional
coupling, and ${\Lambda}_{\rm uv}$ is the ultraviolet cut-off.
\subsubsection*{Perturbative part}
Finally, $Z^{pert}$ is the perturbative
contribution, which is present, since we are working on ${\bR}^4$, and
reduction to quantum mechanics is not valid uniformly everywhere.
Far away
from the origin the gauge fields can be treated perturbatively, and because
of the presence of the rotation ${\exp}{\b}{\Omega}$
one can treat them mode by mode, which gives (cf.  \cite{SmallInst}):
\begin{equation}
Z_5^{\mr{pert}} = \prod_{{\a} \in {\Delta}}
{\Lambda}_5^{\frac{\langle\a, \vec a\rangle^2}{2h^{\vee} {\ve}_1 {\ve}_2}}
\prod_{i,j \geq 0} \left( 1 -
\ex^{{\b}\langle {\a} , {\vec a} \rangle} q_1^i q_2^j \right)
\label{pertur}
\end{equation}
where
\begin{equation}
q_1 = \ex^{{\b}{\ve}_1}, \ q_2 = \ex^{{\b}{\ve}_2}.
\label{qs}
\end{equation}
The infinite product in \Ref{pertur} is absolutely convergent for $|q_1 |,
|q_2 | < 1$, and defines an analytic function in some domain in ${\bC}
\times {\bC}$.
To justify \Ref{pertur} better one introduces
independent ${\bar\Omega}$ so that the mode $f_{i,i',j,j'}z_1^{i-1}{\zb}_1^{i'-1}z_2^{j-1}{\zb}_2^{j'-1}$
of a scalar field $f(z_1 , {\zb}_1 , z_2, {\zb}_2)$
is weighted by $$\ex^{{\b}({\ve}_1 ( i-1) + {\bar\ve}_1 (i'-1) +
{\ve}_2(j-1)+{\bar\ve}_2(j'-1))}$$
In the limit of small ${\b}, {\ve}_1, {\ve}_2$
we can expand $Z^{\mr{pert}}_5$ as follows:
\begin{equation}
\label{pertpr}
Z_5^{\mr{pert}} ( {\vec a}, {\Lambda}_5, {\ve}_1, {\ve}_2 ; {\b} ) \sim
{\exp} \frac{1}{{\ve}_1 {\ve}_2} {\CF}_{\mr{pert}}({\vec a}, {\Lambda}_4) + \ldots
\end{equation}
where
$$
{\Lambda}_5 = ({\b}{\Lambda}_4)^{2 h^{\vee}}
$$
and $\Lambda_4$ is kept finite.
\subsection{Tracing over instanton moduli space}

In this subsection we explain how to calculate the twisted Witten index of
the instanton moduli space, for any gauge group $G$ of type $A,B,C,D$.

The beautiful fact about these moduli spaces which we shall exploit is the
existence of the ADHM construction \cite{ADHM}, which realizes the moduli
space of $G$-instantons with charge $k$ (we shall present it in full detail
below) as the quotient of the space ${\CN}_{G,k}$ of solutions to some quadratic equations
in some vector space ${\CV}_{G,k}$ by the action of some group $G^{D}_k$,
which depends on $G$ and $k$. Here we use the so-called complex description,
in which all the spaces are complex, equations are holomorphic, and the dual
groups complex algebraic. There are different ways to perform the quotient
with respect to $G^{D}_k$ if this group contains center (this is related to the existence
of noncommutative deformation of the instanton equations for $G= U(N)$). All these
differences
will be immaterial.

The group $G \times U(2)$ acts on ${\CV}_{G,k}$ and this action can be
extended to the holomorphic action of $G_{\bC} \times GL(2, {\bC})$.
The equations which cut out ${\CN}_{G,k} \subset {\CV}_{G,k}$ are covariant with respect to
$G_{\bC} \times GL(2, {\bC})$. We want
to calculate the character of the action of this latter group on the space
of
holomorphic functions on ${\CM}_{G,k}$ (in fact, twisted Witten index
coincides with this trace in our case).

We shall now explain how to employ this representation of ${\CM}_{G,k}$.

\noindent\underline{Remark.} In what follows we use the notation:
\begin{equation}
Ch_{\CM} (q) \equiv S{\Tr}_{H({\CM})} (q) \equiv  \sum_{p=0}^{\infty}
(-)^p {\Tr}_{H^{0,p}({\CM})}(q)
\label{notrt}
\end{equation}
\subsubsection{A model example}

First, consider the simpler situation. Suppose that we have a vector space
${\CV} = {\bC}^{K}$, $M$ polynomials $f_1 (z), \ldots, f_{M}(z)$, which
define affine algebraic variety ${\CN} = f^{-1}_{1}(0) \cap \ldots \cap f_{M}^{-1}(0)$.
We assume that they are in generic position, i.e. the matrix $\Vert {\p}_{z_{i}}
f_{l} \Vert$ has maximal rank everywhere except the origin.
We wish to calculate the trace of the action of the torus ${\bT} ({\bC}^{*})^{L} \subset ({\bC}^{*})^{K}$ on the space of holomorphic (polynomial) functions on ${\CN}$,
assuming that the equations $f_l (z) = 0$
are ${\bT}$-equivariant.

Let $q = ( q_1 ,  \ldots , q_L ) \in {\bT}$. Let $w_1(q), \ldots, w_{K}(q)$
be the eigenvalues of the action of $q$ in ${\bC}^K$:
\begin{equation}
w_{i}(q) = \prod_{l=1}^{L} q_{l}^{{\bf w}_{li}}, \qquad {\bf w}_{li} \in {\bZ}
\label{wt}
\end{equation}
Equivariance of the
equations means that
$$
f_{m}( q \cdot z) \equiv
f_{m} ( w_1 (q) z_1 , \ldots w_{K} (q) z_{K} ) = {\vf}_m (q) f_{m} (z)
$$
where ${\vf}_l(q)$ are some monomials in $q$:
\begin{equation}
{\vf}_m (q) = \prod_{l=1}^{L} q_l^{{\bf s}_{lm}}, \qquad {\bf s}_{lm} \in
{\bZ}
\label{ws}
\end{equation}

We start with the space ${\CV}$. Polynomial functions on ${\CV}$
are sums of the monomials. Monomials are eigenspaces of the action of $q$.
The character is, therefore, the sum of all monomials:
\begin{equation}
{\Tr}_{{\bC}^{K}} q = \frac{1}{(1-w_1(q) ) \ldots (1 - w_{K}(q) )}
\label{vect}
\end{equation}
Now, what are the functions on ${\CN}$? Clearly, these are all polynomials
on ${\bC}^K$ modulo those which vanish on ${\CN}$, i.e. modulo polynomials
which can be represented in the form:
\begin{equation}
f_1 (z) m_1 (z) + \ldots f_{M} (z) m_{N} (z)
\label{idel}
\end{equation}
Since $m_1 (z) , \ldots , m_{M}(z)$ are arbitrary polynomials we should
subtract their contribution from \Ref{vect}. Put another way, add:
\begin{equation}
- \frac{{\vf}_1 (q) +  \ldots + {\vf}_{M} (q)}{(1-w_1(q) ) \ldots (1 - w_{K}(q) )}
\label{fstterm}
\end{equation}
where $l$'th term corresponds to contribution of the polynomials of the form
$f_l (z) m_l (z)$.

Have we account for  everything? Not quite. Indeed, there is some redundacy
in \Ref{idel}. Say, we add to $m_{l}(z)$ the polynomial $f_{n}(z) k_{ln}(z)$
and subtract from $m_{n}(z)$ the polynomial $f_{l}(z) k_{ln}(z)$. Clearly,
\Ref{idel} will not change, while we have subtracted the contribution of
such polynomials twice in \Ref{fstterm}. Put another way, the functions
$f_l(z) f_{n}(z) m_{ln}(z)$, for $l \neq n$ are accounted for twice. Thus, we should them it back:
\begin{equation}
+ \sum_{l < n} \frac{{\vf}_{l}(q) {\vf}_{n}(q)}{(1-w_1(q) ) \ldots (1 - w_{K}(q) )}
\label{sndtrm}
\end{equation}
Clearly, we now have the redundacy at the next level, which we should
account for by subtracting triple products etc. Finally we arrive at:
\begin{equation}
Ch_{\CN} (q) = {\Tr}_{Fun({\CN})} ( q) = \frac{\prod_{l=1}^{M} ( 1 -
{\vf}_{l}(q))}{\prod_{i=1}^{K} ( 1 - w_i (q) )}
\label{chr}
\end{equation}
Note that the character is analytic function on ${\bC}^K$ with
the possible poles at $q$'s, such that
some of $w_i(q)$'s are equal to $1$. These come from the noncompact
nature of the space we study. Of course, were ${\CN}$ compact there would be
no holomorphic functions, except for constants. The character \Ref{chr} would be equal to $1$ in this case. The
correct problem in the case of compact ${\CN}$ is to study the
index of ${\bar\p}$ operator coupled to some non-trivial line bundle. In the
context of gauge theories, the compact ${\CN}$ correspond to the moduli
spaces of instantons on compact four dimensional manifolds, and the
non-trivial line bundle comes from observables in the gauge theory.

\noindent\underline{Example}. As an example, which also shows that our formalism
works for singular spaces, let us take the conifold, i.e. the
hypersurface in ${\bC}^4$, given by the equation
$$
f(z) = \sum_{\a = 1}^{4} z_{\a}^2 = 0
$$
In order to exhibith its toric symmetry, write $$X = z_{1} + i z_2, Y = z_{1}
- i z_{2}, Z = z_{3} + i z_{4}, T = z_{3} - i z_{4},$$
then:
$$
f = XY + ZT
$$
it is invariant under the action of the torus ${\bC}^3$:
$$
(q_1, q_2, q_3): (X, Y, Z, T) \mapsto ( q_1 X, q_2 q_3 Y, q_1 q_3 Z, q_2 T )
$$
This action of the torus corresponds to:$$
w_1 (q) = q_1 , w_2(q) = q_2 q_3 , w_3 (q)  = q_1 q_3 , w_4 (q) = q_3 ;
\qquad
{\vf}_1 (q) = q_1 q_2 q_3
$$
By our general formulae we get the character:
\begin{equation}
Ch_{\mathrm{conifold}} (q) = \frac{1 - q_1 q_2 q_3}{(1 - q_1) ( 1 - q_2 ) ( 1- q_1 q_3 )
(1 - q_2 q_3 )}
\label{chrcon}
\end{equation}

Now let us consider more complicated problem. Suppose, in addition, that we
want only functions on ${\CN}$ which are invariant under the linear action of some
group $H_{\bC} \subset GL_{K}({\bC})$ which acts on ${\CV}$, preserving
${\CN}$. On the space $Fun ({\CM})$, for ${\CM} = {\CN}/G_{\bC}$ of functions, invariant under $H_{\bC}$ acts, in
general, a smaller torus ${\bT}^{H} \subset {\bT}$. We want its character.

Clearly, what we should do is to project onto the $H_{\bC}$ -- invariant
functions. Since for holomorphic functions the condition of $H_{\bC}$
invariance is equivalent to the condition of $H$-invariance, where $H$ is
the maximal compact subgroup of $H_{\bC}$ we can perform the projection by
the simple integral over $H$:
$$
Ch_{{\CN}/H_{\bC}}(q) = \frac{1}{{\rm Vol}(H)} \int_{H} \dd h\ Ch_{\CN} ( q \cdot h
)
$$
where $dh$ is the Haar measure on $H$ and ${\rm Vol}(H)  = \int_{H} \dd h$.
Here $q \in {\bT}^{H}$. Since the trace depends only on the conjugacy class,
we may assume that $h$ is in the maximal torus ${\bT}_{H} \subset H$ of the compact subgroup.
We may also assume that ${\bT}_{H} \subset {\bT}$, by choosing appropriately the basis in
${\CV}$.
Together, $q \cdot h$ are in ${\bT}$. Its trace on $Fun({\CN})$ we already
know from \Ref{chr}. What remains is to integrate over $H$, i.e. over
${\bT}_{H}$ and then include the volume of the adjoint orbit of $h$, which
is the famous Weyl-Vandermonde factor:
\begin{equation}
Ch_{{\CM}} (q) = \frac{1}{\vert W \vert {\rm Vol}{\bT}_{H}} \int_{{\bT}_{H}} \dd h
\
\prod_{{\a} \in {\Delta}} \left( h^{{\a}/2} - h^{-{\a}/2} \right)
\
\frac{\prod_{l=1}^{M} ( 1 -
{\vf}_{l}(q h))}{\prod_{i=1}^{K} ( 1 - w_i(q h) )}
\label{projchr}
\end{equation}
{\em Note that the integral in \Ref{projchr} is well-defined, as it is an
integral over compact group, and the integrand has no poles, as long as we
keep $\rm q$ outside of compact torus.}

\noindent\underline{Example.} Let us go back to our conifold example, but
now let us view it as the quotient of ${\bC}^4$ by the action of ${\bC}^*$:
$$
(z_1, z_2, z_3, z_4) \mapsto ( wz_1 , wz_2 , w^{-1} z_3, w^{-1} z_4)
$$
The torus $({\bC}^*)^3$ acts as follows:
$$
(z_1, z_2, z_3, z_4) \mapsto ( q_1 z_1 , q_2 z_2 , q_3  z_3,  z_4)
$$
The coordinates on the quotient space are:
$$
X = z_1 z_4, \ Y = z_2 z_3, \ Z = z_1 z_3, \ T = - z_2 z_4
$$
which obey $X Y + Z T = 0$, as in our previous definition of the conifold.
Now let us calculate the character, using \Ref{projchr}
\begin{equation}
Ch_{\mathrm{conifold}} (q) = \oint \frac{\dd z}{2\pi i \ z} \frac{1}{(1 - q_1 z)(1 - q_2
z )(1 - q_3 z^{-1})( 1 -z^{-1} )} \label{chrconii}
\end{equation}
$$
\frac{1}{( 1 - q_1 q_3 ) ( 1 - q_2 q_3 ) ( 1 - q_3^{-1} ) } + \frac{1}{(1-q_1 )(1 - q_2 ) ( 1- q_3 ) }
$$
which coincides with \Ref{chrcon}. The expression which we just got also has
another meaning. The two terms above come from two fixed points of the torus
action on the {\it resolved conifold}, which is the
total space of the bundle ${\CO}(-1) \oplus {\CO}(-1)$ over ${\bC\bP}^1$.
However, our character does not depend on the kahler moduli of the
resolution, and makes sense for the singular space as well.

\subsubsection{Back to instantons}

Now let us review the problem at hands.
We wish to calculate the character
of the moduli space of instantons, ${\CM}_{G,k}$. For each value of the
instanton charge this space is the quotient of the form we discussed above.
Let ${\bT}_k$ denote the maximal torus of $G^{D}_{k}$, $r_k$ its rank,
$W_k$ its Weyl group.
Then:
\begin{multline}
Ch_{{\CM}_{G,k}} (q) = \frac{1}{|W_{k}| {\rm Vol}{\bT}_{k}}
\oint_{{\bT}_k} \prod_{i=1}^{r_k} \dd{\phi}_i \prod_{{\a} \in {\Delta}^{+}}
\left\vert \sin \frac{\langle {\a}, {\phi} \rangle}{2} \right\vert^2 \\
\frac{\prod ( 1 - {\vf}_{\mathrm{ADHM \ equations}} (q, \ex^{i {\phi}}))}
{\prod ( 1 - {w}_{\mathrm{ADHM \ matrices}} (q, \ex^{i {\phi}}))}
\label{instchr}
\end{multline}
Here $q = ( \ex^{{\b}{\vec a}}, \ex^{{\b}{\Omega}} )$. In the formula \Ref{instchr} we
have the ratio of terms, the denominator comes from the weights of the
{\em complex} ADHM matrices, and the numerator comes from the weights
of the {\em complex} ADHM equations ${\m}_{\bC}$, which we describe in
detail below. ADHM equations also contain the so-called real equation
${\m}_{\bR}$.
Its contribution is, in some sense,
the Weyl-Vandermonde factor.

Thus, \Ref{instchr} together with \Ref{pertur}
gives us all the necessary ingredients for \Ref{part}.

\subsection{Four dimensional limit}

In this paper we shall not explore five dimensional theory in full
generality. We want to take the limit ${\b} \to 0$, while scaling
${\Lambda}_5$ in such a way that the instanton effects are finite.
In the integral \Ref{projchr} this limit corresponds to taking $ q \ex^{{\b}{\bX}}, h = \ex^{{\b}{\phi}}$, ${\phi} \in
Lie(G_{k}^{D})$, and sending ${\b} \to 0$ while keeping ${\bX},
{\phi}$ finite. In this limit the integral over ${\phi}$
becomes noncompact
one, over the Lie algebra of ${\bT}_{H}$. We can view it as a contour
integral of a meromorphic top degree form over complexified Lie algebra. The integrals
\Ref{projchr},
\Ref{instchr}
will scale as
the $K - M - \dim(H) = \dim{\CM}$-th power of ${\b}$:
\begin{equation}
Ch_{{\CM}} \ex^{{\b}{\bX}} \sim {1\over{{\b}^{{\dim}{\CM}}}} \frac{1}{
\vert W_{H} \vert {\rm
Vol}({\bT}_{H})}\int_{Lie({\bT}_{H})} {\rm d}{\phi} \ \prod_{{\a} \in {\Delta}} \langle {\phi}, {\a} \rangle \
\frac{\prod_{l=1}^{M}  {\bf s}_{l}( {\bX} \oplus {\phi} )}{\prod_{i=1}^{K} {\bf w}_{i}({\bX} \oplus {\phi})}
\label{projrat}
\end{equation}
where ${\bf w}$, ${\bf s}$ are linear functions on the Lie algebras of
${\bT}_{H}, {\bT}^{H}$, constructed  out of ${\bf s}_{lm}, {\bf
w}_{li}$.

\noindent{\em Note that the integrand in \Ref{projrat} has no poles if ${\phi}$ is
kept on the real slice while ${\bX}$ has imaginary part. The sign of this
part is determined by the convergence of the original five dimensional
character. In the context of gauge theory below this will be the condition
that $q_s = \ex^{{\b}{\ve}_s}$, $s = 1,2$ are inside the unit disk, so that
${\ve}_s$'s have positive imaginary part.}

Our goal is to perform integrals, similar to \Ref{projrat} in the case where ${\CM}$ is the moduli space of instantons.
In order to do so we describe the moduli space of
instantons for $SO(N)$ and $Sp(N)$ in the next section.


\section{ADHM construction}
\label{ADHM}

In this section we present the ADHM construction of instantons  \cite{ADHM}
for all classical groups, namely for $SU(N)$, with algebra $A_{N-1}$, $SO(2N+1)$ with algebra $B_N$, $Sp(N)$,
with algebra $C_N$ and $SO(2N)$, with algebra  $D_N$.

There exist
several nice descriptions of
the construction of instantons for the $SU(N)$
case  \cite{MultiInstCalc,CalcManyInst,InstAndRec}.
We give below the version of
the original work  \cite{SelfDualSolution}, adapted for our purposes.

As  self-duality equation can have real solutions
only in Euclidean space-time, we switch to Euclidean signature in what
follows.


\subsection{$SU(N)$ case}
\label{ADHMSU}

To obtain the solution of the self-dual equation with the instanton number $k$ we need the following ingredients: the $k \times N + 2k$ matrix which depends linearly on the coordinate $x_{\alpha\dot{\alpha}} = x^\mu \sigma_{\mu,\alpha\dot{\alpha}}$:
$$\Delta_{\dot{\alpha}}(x) = \A_{\dot{\alpha}} + \B^\alpha x_{\alpha\dot{\alpha}}$$
satisfying the factorization condition
\begin{equation}
\label{factorization}
\Delta^{\dag\dot{\alpha}}\Delta_{\dot{\beta}} = \R^{-1}\delta^{\dot{\alpha}}_{\dot{\beta}}
\end{equation}
with $\R$ being $k \times k$ invertible matrix. By definition it is hermitian. Also we need $N \times N + 2k$ matrix $v$ satisfying the conditions:
\begin{equation}
\label{vCondition}
\begin{aligned}
v^\dag v &= \I_N  & \  \ &\mbox{and}\ \ & \Delta^{\dag\dot{\alpha}} v &= 0.
\end{aligned}
\end{equation}
Then the connection satisfying the anti-self-duality equation is
$$A_\mu = v^\dag \pd_\mu v.$$

Neither \Ref{factorization} nor \Ref{vCondition} changes under the
transformation
\begin{equation}
\label{transform}
\begin{aligned}
\Delta_{\dot{\alpha}} &\mapsto \Delta_{\dot{\alpha}}' U \Delta_{\dot{\alpha}} M & \ \ &\mbox{and} \ \
& v &\mapsto v' = Uv
\end{aligned}
\end{equation}
with $U$ being a $(N + 2k) \times (N + 2k)$ unitary matrix and $M$ being an invertible one. This freedom can be used to put the matrix $\B = (\B^1, \B^2)$ into the canonical form
$$
\B = \left(
\begin{array}{c}
0 \\
\I_k \otimes \I_2
\end{array}
\right).
$$
Having fixed this form we still have a freedom to perform a transformation \Ref{transform} which for the matrix
$$\A = (\A_{\dot{1}}, \A_{\dot{2}}) = \left(
\begin{array}{c}
S_{\dot{1}}\;\; S_{\dot{2}} \\
X^\mu \otimes \sigma_\mu
\end{array}
\right)
$$
can be read as
\begin{equation}
\label{residual}
\begin{aligned}
S_{\dot{\alpha}} &\mapsto
S_{\dot{\alpha}}' = U_N S_{\dot{\alpha}} U_k^{-1} & \ \ &\mbox{and}\ \
& X^\mu &\mapsto {X^\mu}' = U_k X^\mu U^{-1}_k
\end{aligned}
\end{equation}
where $U_k \in U(k)$ and $U_N \in U(N)$. Matrices $S_{\dot{\alpha}}$ transform under the space-time rotations as righthand spinor and $X^\mu$ --- as vector.

The factorization condition \Ref{factorization} requires the matrices $X^\mu$ to be hermitian: ${X^\mu}^\dag = X^\mu$ and also the following non-linear conditions to be satisfied:
$$\mu^i = \A^{\dag\dot{\alpha}} \tau_{i,\dot{\alpha}}{}^{\dot{\beta}}\A_{\dot{\beta}} = 0.$$

These conditions are known as the ADHM equations. They
are usually written in slightly different notations.
Namely, let
$$
\begin{aligned}
J &= S_{\dot{1}}, \ & I &= {S_{\dot{2}}}^\dag, \ & B_1 &= X^0 - i X^3 & \ \
&\mbox{and}\ \ &  B_2 &= -iX^1 - X^2.
\end{aligned}
$$
Then the ADHM equation are
$$
\begin{aligned}
\mu_\Real &= - \mu^3 = I I^\dag - J^\dag J  + [B_1,B_1^\dag] + [B_2,B_2^\dag] = 0 & \ \
&\mbox{and} \\
\mu_\Compl &= \frac{1}{2}(\mu^1 -i \mu^2) = \mu^- =  IJ + [B_1,B_2] = 0. & &
\end{aligned}
$$
If we consider two vector spaces $\CV_k = \Compl^k$ and $\CW_N = \Compl^N$ then $I, J, B_1$ and $B_2$ become linear operators acting as
$$
\begin{aligned}
I: \CW_N &\to \CV_k,& \ J: \CV_k &\to \CW_N, & \ \  &\mbox{and} \ \ & B_1, B_2: \CV_k &\to \CV_k.
\end{aligned}
$$

The space of such operators modulo transformations \Ref{residual} is the instanton moduli space. The former statement can be proven by direct calculation using the definition \Ref{InstNumb} and the Osborn's formula  \cite{OsbornIdentity}
\begin{equation}
\label{Osborn}
\Tr_{\rm{fund}} (F_{\mu\nu} \star F^{\mu\nu}) = -\square^2 \ln \det \R
\end{equation}
where the trace in the righthand side is taken over defining (fundamental) representation of $SU(N)$.

The residual freedom \Ref{residual} corresponds to the freedom of the framing change in $\CV_k$ and $\CW_N$. Framing change in $\CW_N$ corresponds to the rigid gauge transformation, which change, in particular, the gauge at infinity, whereas the framing change in $\CV_k$ becomes natural when one considers the instanton moduli space as a hyper-K\"ahler quotient.

Indeed, the space of all (unconstrained) matrices
 $\A_{\dot{\alpha}}$ has a
natural metric $\dd\A^{\dag\dot{\alpha}}\dd\A_{\dot{\alpha}}$
and the hyper-K\"ahler structure
which consists of the triplet of linear operators $\mathcal{I}^i$
which together with the identity operator is
isomorphic to the quaternion algebra.
These operators act as follows:
$$\mathcal{I}^i \A_{\dot{\alpha}} = -i \tau_{i,\dot{\alpha}}{}^{\dot{\beta}}\A_{\dot{\beta}}.$$
The action of the unitary group $U(k)$ described by \Ref{residual} is hamiltonian with respect to each symplectic structure. The Hamiltonian (moment), corresponding to the $i$-th symplectic form and the  algebra element $\xi \in \alg{u}(k)$ is
$$\mu^i_\xi = -i \tau_{i,\dot{\alpha}}{}^{\beta} \Tr (\xi \A^{\dag\dot{\alpha}}\A_{\dot{\beta}}).$$
Hence the ADHM equations together with residual transformation can be
interpreted as the hyper-K\"ahler quotient  \cite{Rocek}:
$${\CM}_{G,k} = \mu^{-1}(0) / U(k).$$

We call the group which is responsible to the framing change in $\CV_k$ the dual group.
 In the case of $SU(N)$
the dual group is $U(k)$.


\subsection{$SO(N)$ case}

The extension to the $SO(N)$ case can be obtained with the help of the reciprocity construction  \cite{InstAndRec}. Consider the Weyl equation for the spinors $\psi^\alpha$ in the fundamental representation of $SU(N)$:
$$\D^{\dag\dot{\alpha}\alpha}\psi_\alpha = \sigma^{\dag}_\mu{}^{\dot{\alpha}\alpha}\D_\mu^\dag \psi_\alpha = 0$$
where $\D_\mu = \pd_\mu + A_\mu$ is the covariant derivative. Its $k$ independent solutions can be arranged to the $N \times k$ matrix $\psi^\alpha = v^\dag \B^\alpha \R$  \cite{OsbornDirac}. One can show that the following statements hold:
\begin{equation}
\label{reciprocity}
\begin{aligned}
\int \dd^4x \psi^{\dag}_\alpha \psi^\alpha & = \pi^2 \I_k, & \int \dd^4x \psi^{\dag}_\alpha\psi^\alpha x^\mu &= -\pi^2 X^\mu & \ \
&\mbox{and} \\
\psi^\alpha x_{\alpha\dot{\alpha}} &\to - \frac{1}{x^2} S_{\dot{\alpha}} & \mbox{when}\;\;\;\; x &\to \infty.
\end{aligned}
\end{equation}
Taking these equations as the definitions of $X^\mu$ and $S_{\dot{\alpha}}$ one recovers both  the ADHM constraints and the fact that the matrices $X^\mu$ are hermitian.

Since for the simple groups the Killing metric is unique up to multiplicative factor we conclude that for all representations $\Tr_{\mr{adj}}(F \wedge F) = \ell_{\mr{rep}} \Tr_{\mr{rep}} (F \wedge F)$ where $\ell_{\mr{rep}}$ is the (Dynkin) index of this representation. For the defining (fundamental) representations of $SU(N)$, $SO(N)$ and $Sp(N)$ we have the following index values:
\begin{equation}
\label{indices}
\begin{aligned}
\ell_{SU(N)} &= 2h^\vee_{SU(N)} = 2N, \\
\ell_{SO(N)} &= h^\vee_{SO(N)}  = N - 2, \\
\ell_{Sp(N)} &= 2h^\vee_{Sp(N)} = 2(N + 1).
\end{aligned}
\end{equation}
Therefore formula \Ref{InstNumb} together with \Ref{Osborn} shows that in the case of $SO(N)$ to obtain the solution of the self-dual equation with the instanton number $k$ we should replace $k$ by $2k$ in the construction for $SU(N)$.

Let us split the index which runs over $1,\dots,2k$ into two indices: one of them, which we denote $A$, will run over $1,2$, and the other --- over $1,\dots,k$. Thus the solution of the Weyl equation can be written as the set of four $N \times k$ matrices $\psi_{\alpha A}$.
These matrices can be represented as follows:
$$\psi_{\alpha A} = \psi_\mu\sigma^\mu_{\alpha A}.$$
The twisted index $\mu$ that appears in the righthand side does not correspond to a Lorentz vector. The Weyl equation can be rewritten now as a set of four equations:
\begin{equation}
\label{psiEqn}
\begin{aligned}
\D^\mu \psi_\mu &=0 & \ \ &\mbox{and} \ \ & {(\D_\mu\psi_\nu)}^- = 0
\end{aligned}
\end{equation}
where ${(F_{\mu\nu})}^- = \frac{1}{2}(F_{\mu\nu} - \star F_{\mu\nu})$ is the anti-selfdual part of an antisymmetric tensor $F_{\mu\nu}$. It worth noting that these conditions mean that $\psi_\mu$ is orthogonal to the gauge transformations and that it satisfies the linearized self-dual equation.

The condition that $A$ belongs
to the algebra of $SO(N)$ implies that $A_{\m}$ are real skew-symmetric matrices.
Hence the equation for $\psi_\mu$ has
real coefficients and its solutions can be chosen real as well.
The fact that $\psi_\mu$ are real means that
$\psi_\mu \sigma^\mu$ can be considered as a quaternion.
We recover here the quaternion construction introduced
 in  \cite{SelfDualSolution}.
The possibility of this expansion with
real coefficients implies that $S_{\dot{\alpha}}$ can also be expanded as
$S_{A\dot{\alpha}} = S_\mu \sigma^\mu_{A\dot{\alpha}}$
where $S_\mu$ are real.

Using then the definition of $X^\mu$ \Ref{reciprocity} we derive the following statement:
$$\epsilon_{CA} X^{\mu,A}{}_B \epsilon^{BD} =  {({X^\mu}^\T)}_C{}^D,$$
or, if we introduce the symplectic structure ${\bJ}_{2k}$
this can be written as
$${\bJ}_{2k} X^\mu {\bJ}_{2k}^\T = {X^\mu}^\T.$$

The dual group is a subgroup of $U(2k)$ which preserves this condition. It is the group $Sp(k) \subset U(2k)$.

The matrices $X^\mu$ and $S_{\dot{\alpha}}$
can be represented as follows:
\begin{equation}
\label{SOYZIJ}
\begin{aligned}
X^\mu &= \left(
\begin{array}{cc}
Y^\mu & {Z^\mu}^\dag \\
Z^\mu & {Y^\mu}^\T
\end{array}
\right),\quad
& \quad S_{\dot{1}} = J = (K,K') &
& \ \ &\mbox{and}\ \  &
S_{\dot{2}} &= I^\dag = (-K^{' \ast},K^{\ast})
\end{aligned}
\end{equation}
where $Y^\mu$ is an hermitian matrix and $Z^\mu$ is an antisymmetric one.

Let
\begin{equation}
\label{SOBB}
B_{1,2} = \left(
\begin{array}{cc}
P_{1,2} & Q_{1,2}' \\
Q_{1,2} & P_{1,2}^\T
\end{array}
\right),
\end{equation}
where $Q_{1,2}$ are antisymmetric matrices. The ADHM equations for $SO(N)$ becomes:
\begin{equation}
\label{SOMN}
\begin{aligned}
\mu_\Compl &= \left(
\begin{array}{cc}
M_\Compl & N_\Compl' \\
N_\Compl & -M_\Compl^\T
\end{array}
\right) = 0 & \ \
&\mbox{and} \ \ &
\mu_\Real &= \left(
\begin{array}{cc}
M_\Real & N_\Real' \\
N_\Real & -M_\Real^\T
\end{array}
\right) = 0
\end{aligned}
\end{equation}
where
$$
\begin{aligned}
M_\Compl &= [P_1,P_2] + Q'_1 Q_2 - Q'_2 Q_1 - K^{' \T} K, \\
N_\Compl &= Q_1 P_2 - P_2^\T Q_1 + P_1^\T Q_2 - Q_2 P_1 + K^\T K,\\
N_\Compl' &=  Q_1' P_2^\T- P_2 Q_1' + P_1 Q_2'- Q_2' P_1^\T  - K^{' \T} K',
\end{aligned}
$$
and
$$
\begin{aligned}
M_\Real &= \sum_{s = 1}^2 \left([P_s,{P_s}^\dag] + {Q_s}^\ast Q_s - {Q_s}' {Q_s}^{' \ast} \right) + K^{' \T} K^{' \ast} -
K^\dag K, \\
N_\Real &= \sum_{s = 1}^2 \left(Q_s {P_s}^\dag - {P_s}^\ast Q_s  + {Q_s}^{' \ast} P_s- {P_s}^\T {Q_s}^{' \ast}
\right)- K^\T K^{' \ast} - K^{' \dag} K, \\
N_\Real' & = \sum_{s = 1}^2 \left( {Q_s}' {P_s}^\ast - {P_s}^\dag {Q_s}' + {Q_s}^\ast {P_s}^\T - P_s {Q_s}^\ast \right) - K^{' \T} K^\ast - K^\dag K'.
\end{aligned}
$$
Note that $N_\Compl, N'_\Compl, N_\Real$ and $N_\Real'$ are symmetric matrices.


\subsection{$Sp(N)$ case}

The group $Sp(N)$ is a subgroup of $SU(2N)$ which
preserves the symplectic structure ${\bJ}_{2N}$.
The ADHM construction for $Sp(N)$ can be obtained
by imposing some constraints on the ADHM construction for $SU(2N)$.
A quick look at \Ref{indices}
shows that in this case there is no  doubling of the instanton charge.

Let us choose the Darboux basis in ${\bC}^{2N}$, which corresponds to the
split ${\bC}^{2N} = {\bC}^2 \otimes {\bC}^N$, ${\bJ}_{2N} = {\bJ}_2 \otimes
{\bI}_{N}$.
Correspondingly, we
split the index which runs over $1,\dots,2N$ into two:
the first, $A= 1,2$, and the second: $l = 1,\dots,N$.

We can expand the solution of the
Weyl equation as follows $\psi_{\alpha}^{A,l} = \psi_{\mu}^l
\sigma^\mu_{\alpha A}$.
The fact that $A_\mu$ belongs to $\alg{sp}(N)$
imposes the following condition:
$${\bJ}_{2N} A_\mu^\ast {\bJ}_{2N} = A_\mu.$$
The solutions $\psi_{\mu}^l$
can be chosen to be real. Thus the reciprocity formulae \Ref{reciprocity} show that in that case the matrices $X^\mu$ are not only hermitian, but also real and consequently symmetric. The dual group should preserve this condition and we arrive to the conclusion that this is $O(k) \subset U(k)$.

The reality of $\psi_{\mu}^l$ implies also that the matrices $S_{\dot{\alpha}}$ can be expanded as $S_{A\dot{\alpha}} = S_\mu \sigma^\mu_{A \dot{\alpha}}$ where $S_\mu$ are real. Hence for the matrices $I$ and $J$ we have
\begin{equation}
\label{IJSpN}
\begin{aligned}
J &= \left(
\begin{array}{c}
K \\
K'
\end{array}
\right) & \ \ &\mbox{and}\ \ & I^\dag &= \left(
\begin{array}{c}
- K^{' \ast} \\
K^\ast
\end{array}
\right).
\end{aligned}
\end{equation}

Hence the ADHM equation for $Sp(N)$ take the following form
$$
\begin{aligned}
\mu_\Compl &= K^\T K' - K^{' \T} K + [B_1,B_2], \\
\mu_\Real &= K^\T K^\ast - K^\dag K + K^{' \T} K^{' \ast} - K^{' \dag} K' + [B_1,B_1^\ast ] + [B_2,B_2^\ast].
\end{aligned}
$$
Here the matrices $B_{1,2}$ are symmetric. We see that $\mu_\Compl$ and $\mu_\Real$ are antisymmetric matrices.


\section{Derivation of the prepotential}
\label{Derivation}

In this section we derive
the prepotential for the $SO(N)$ and $Sp(N)$
cases as a formal series and after,
applying the variational technique of  \cite{SWandRP}
we obtain the Seiberg-Witten data.

\subsection{Five dimensional expression}

We start with five dimensional instanton partition functions.
>From ADHM constructions  \cite{ADHM} we have all the ingredients:
the vector spaces ${\CV}_{k}$, the equations ${\m}_{\bC} = 0$, which
define ${\CN}_k$, and the dual groups $G_{k}^D$.

\subsubsection{The Haar measures}

In what follows we need the explicit expressions for the Haar measures on
the
dual groups, pushed down to their maximal tori.

The general formula for the Haar measure reduced to the maximal
torus of the group is given by ($r$ being the rank of the group):
$$
\dd\mu_G = \frac{1}{|W|} \prod_{i=1}^r \frac{\dd\phi_i}{2\pi} {|{\d}|}^2
$$
where
$$
\d = \prod_{\alpha \in {\Delta}^+} \left(
\ex^{\frac{i}{2}\langle \a, \phi\rangle} - \ex^{-\frac{i}{2}\langle \a, \phi\rangle}\right),
$$
This measure gives a measure on the Lie algebra
(it corresponds to the limit of  small $\phi_i$)
$$
\dd\mu_{\alg{g}} = \frac{1}{|W|} \prod_{i=1}^k \frac{\dd\phi_i}{2\pi} \prod_{\alpha \in
{\Delta}^+} {\langle\alpha, \phi\rangle}^2 .
$$

Using the root systems of algebras of type $B$, $C$ and $D$ we can write the measures:
$$
\begin{aligned}
\dd\mu_{B_k} &= \frac{1}{2^k k!} \prod_{i=1}^k \frac{\dd\phi_i}{2\pi} \prod_{i<j}{(\phi_i^2 - \phi_j^2)}^2 \prod_{i=1}^k \phi_i^2, \\
\dd\mu_{C_k} &= \frac{1}{2^k k!} \prod_{i=1}^k \frac{\dd\phi_i}{2\pi} \prod_{i<j}{(\phi_i^2 - \phi_j^2)}^2 \prod_{i=1}^k (2 \phi_i)^2, \\
\dd\mu_{D_k} &= \frac{1}{2^{k-1} k!} \prod_{i=1}^k \frac{\dd\phi_i}{2\pi} \prod_{i<j}{(\phi_i^2 - \phi_j^2)}^2 .
\end{aligned}
$$
The detailed investigation of the Haar measure can be found for example in  \cite{Bourbaki}, and in \cite{KrStau}.

\subsubsection*{The case of $C_{N}$}

Consider first the $Sp(N)$ case. Let $k = 2 n  +{\chi}$, ${\chi} = 0,1$.
Let us choose the matrices from the Cartan subalgebra of
$Sp(N)$ and $SO(k)$ in the standard forms:
$$
\begin{aligned}
a &= \left(
\begin{array}{cc}
\tilde{a} & 0 \\
0 & -\tilde{a}
\end{array}
\right)  \in \alg{sp}(N) \;\;\;\mbox{where}\;\;\; \tilde{a} = \diag \left\{ i a_1,\dots,i a_N\right\}, \\
\phi &= \diag \left\{
\left(
\begin{array}{cc}
0 & -\phi_1 \\
+\phi_1 & 0
\end{array}
\right),\dots,
\left(
\begin{array}{cc}
0 & -\phi_n \\
+\phi_n & 0
\end{array}
\right), \diamondsuit \right\} \in \alg{so}(k).
\end{aligned}
$$
Here $\diamondsuit = 0$ for odd $k$ and is absent for even $k$.
The eigenvalues $a_1,\dots,a_N$ and $\phi_1,\dots,\phi_n$ are assumed real.
Let $z_i = \ex^{i{\b} {\phi}_i}$, $t_l = \ex^{i {\b}a_l}$, $q_s \ex^{{\b}{\ve}_s}$, $s=1,2$. Our conventions imply that $| z_i | = 1$, $| t_l
| = 1$, $| q_s | < 1$.
Introduce:
$$
\begin{aligned}
{\Delta}_{\b}(q) &= \prod_{i=1}^{n} (  1 - q z_i )^{\chi} ( 1  - q /z_{i} )^{\chi}\\
&\times \prod_{i < j} ( 1 - q z_i z_j ) ( 1 - q z_i /z_j )( 1 - q z_j / z_i )( 1 - q / (z_i z_j)
), \\
{\CP}_{\b} (y) &=  \prod_{l=1}^{N}  ( 1 - t_l y ) ( 1 - y/t_l).
\end{aligned}
$$
Then we have:
\begin{multline}
\label{Zspb}
Z_k (a, {\b};{\ve}_1,{\ve}_2) = \frac{1}{2} \frac{{(-1)}^{{\chi}}}{2^{n-1+{\chi}} n!}
\frac{(1 - q_1 q_2)^n}{(1-q_1)^{n+{\chi}}(1-q_2)^{n+{\chi}} {\CP}_{\b}((q_1
q_2)^{\frac{1}{2}})^{\chi}}
 \\
\times \oint \prod_{j=1}^n \frac{\dd z_{j}}{2\pi i z_j} \frac{\Delta_{\b}(1)\Delta_{\b}(q_1 q_2)}{
\Delta_{\b}(q_1)\Delta_{\b}(q_2)}  \\
\times
\prod_{j=1}^{n} {\left[  \Pol_{\b}(z_j (q_1 q_2)^{-\frac{1}{2}})
\Pol_{\b}(z_{j} (q_1 q_2)^{\frac{1}{2}})
\prod_{s=1}^2  \left\{ ( 1 - q_s z_j^2) ( 1 - q_s / z_j^2 ) \right\}
\right]}^{-1}
\end{multline}
The factor $2^{n-1 + \chi}n!$ outside of the integral in \Ref{Zspb}
is the order of the Weyl group of the dual group, which is $O(k)$.
The factor $\dfrac{1}{2}$ is the ratio of the volumes of the dual group and
its connected subgroup $SO(k)$.

\subsubsection{The case of $B_n$ and $D_n$}

Now we take $G = SO(N)$.
This time, let $N = 2n +{\chi}$, ${\chi} = 0,1$. The case $\chi = 0$
corresponds to $D_n$ and $\chi =1 $ gives $B_n$.

As in the previous section we choose the matrices
from Cartan subalgebras of $G$ and $G^D_k$ in the standard form:
$$
\begin{aligned}
a &= \diag \left\{  \left(
\begin{array}{cc}
0 & -a_1 \\
+a_1 & 0
\end{array}
\right),\dots,\left(
\begin{array}{cc}
0 & -a_n \\
+a_n & 0
\end{array} \right),\diamondsuit \right\} \in \alg{so}(N), \\
\phi &= \left(
\begin{array}{cc}
\tilde{\phi} & 0 \\
0 & -\tilde{\phi}
\end{array} \right) \in \alg{sp}(k)
\;\;\;\mbox{where}\;\;\; \tilde{\phi}  = \diag\{i\phi_1,\dots,i\phi_k\}.
\end{aligned}
$$
Here $\diamondsuit = 0$ for odd $N$ and is absent for even $N$.  Here, as before,
$\phi_1,\dots\phi_k$ and $a_1,\dots,a_n$ are supposed to be real. Again, we
introduce: $z_i = \ex^{i{\b} {\phi}_i}$, $i = 1, \ldots, k$,
$t_l = \ex^{i {\b}a_l}$, $l = 1, \ldots , n$, $q_s \ex^{{\b}{\ve}_s}$, $s=1,2$. Our conventions imply that $| z_i | = 1$, $| t_l
| = 1$, $| q_s | < 1$.

Introduce:
$$
\begin{aligned}
{\Delta}_{\b}(q) &\prod_{i < j} ( 1 - q z_i z_j ) ( 1 - q z_i /z_j )( 1 - q z_j / z_i )( 1 - q / (z_i z_j)
),\\
{\CP}_{\b} (y) &=  ( 1 - y )^{\chi} \prod_{l=1}^{n}( 1 - t_l y ) ( 1 - y/t_l)
\end{aligned}
$$
Then we can write:
\begin{multline}
\label{ZSOb}
Z_k (a, {\b};{\ve}_1,{\ve}_2) = \frac{(-)^{kN}}{2^{k} k!}
\frac{(1 - q_1 q_2)^k}{(1-q_1)^{k}(1-q_2)^{k}}
 \\
\times \oint \prod_{j=1}^n \frac{\dd z_{j}}{2\pi i z_j}
\frac{\Delta_{\b}(1)\Delta_{\b}(q_1 q_2)}{
\Delta_{\b}(q_1)\Delta_{\b}(q_2)}  \\
\times
\prod_{j=1}^{k} \frac{( 1 - z_j ) ( 1- 1/z_j)(1 - (q_1 q_2)^{\frac{1}{2}}z_j)
(1 - (q_1 q_2)^{\frac{1}{2}}/z_j)}{\Pol_{\b}(z_j (q_1 q_2)^{-\frac{1}{2}})
\Pol_{\b}(z_{j} (q_1 q_2)^{\frac{1}{2}})}
\end{multline}


\subsection{Four dimensional limit}

We now take the four dimensional limit, ${\b} \to 0$.
\subsubsection*{$Sp(N)$ case}
In the limit ${\b} \to 0$ we have (set $q = \ex^{i{\b}t}$, $ y = \ex^{i{\b}x}$):
$$
\begin{aligned}
{\CP}_{\b} &\to {\b}^{2N} {\CP}(x), &\qquad
{\Delta}_{\b}(q) &\to {\b}^{2 n {\chi} + 2n (n-1)} {\Delta}(t),
\end{aligned}
$$
where
$$
\begin{aligned}
{\Delta}(t) &= \prod_{i=1}^{n} (  {\phi}_i ^2  - t^2 )^{\chi}
\prod_{i < j} (({\phi}_i + {\phi}_j  )^2 - t^2) (({\phi}_i - {\phi}_j  )^2 - t^2),
\\
\qquad {\CP} (x) &=  \prod_{l=1}^{N}  ( x^2 - a_l^2 ).
\end{aligned}
$$
Then we have
\begin{equation}
\begin{aligned}
\label{Zsp}
Z_k(a, {\b};\eps_1,\eps_2) &\to   {\b}^{-2k (N+1)}Z_k(a;\eps_1,\eps_2) , \mbox{ where } \\
Z_k(a;\eps_1,\eps_2) &= \frac{1}{2}\frac{{(-1)}^n}{2^{n-1} n!}\frac{{(\eps_1 + \eps_2)}^n}{{(\eps_1\eps_2)}^n}
{\left[-\frac{1}{2\eps_1 \eps_2  \prod_{l=1}^N ((\eps / 2)^2 - a_l^2)}\right]}^\chi \\
\times &\int \prod_{i=1}^n \frac{\dd\phi_{i}}{2\pi i} \frac{\Delta(0)\Delta(\eps_1 + \eps_2)}{\Delta(\eps_1)\Delta(\eps_2)}  \\
\times
&{\left[\left\{\prod_{s=1}^2 \prod_{i=1}^n \left({(2\phi_{i})}^2 - \eps_s^2\right) \right\}\prod_{j=1}^n \Pol(\phi_{j} - \eps/2) \Pol(\phi_{j} + \eps/2)\right]}^{-1}
\end{aligned}
\end{equation}
Thus we need to scale:
$$
{\Lambda}_5 \propto {\b}^{2(N+1)}{\Lambda}_4^{2(N+1)} = ( {\b} {\Lambda}_4)^{2h^{\vee}_{Sp(N)}}
$$
where ${\Lambda}_4$ is finite, four dimensional QCD scale.

\subsubsection*{$SO(N)$ case}

Similarly, in the limit ${\b} \to 0$ we have:
$$
\begin{aligned}
\Delta_{\b}(q) &\to {\b}^{2k(k-1)} {\Delta}(t), & \qquad {\CP}_{\b} (y) &\to {\b}^{N} {\CP}(x),
\end{aligned}
$$
$$
\begin{aligned}
\Delta(t) &= \prod_{i<j} \left( {\left(\phi_{i} - \phi_{j}\right)}^2 - t^2 \right) \left( {\left( \phi_{i} + \phi_{j} \right) }^2 - t^2 \right), \\
\Pol(x) &= x^\chi\prod_{l = 1}^n \left(x^2 - a_l^2 \right).
\end{aligned}
$$
The instanton integral scales as:
$$Z_k(a, {\b};\eps_1,\eps_2) \to   {\b}^{-2k (N-2)}
Z_k(a;\eps_1,\eps_2)$$
Then we can write:
\begin{equation}
\label{ZSO}
\begin{aligned}
Z_k(a;\eps_1,\eps_2) &= \frac{{(-1)}^{k(N+1)}}{2^k k!} \frac{{(\eps_1 + \eps_2)}^k}{{(\eps_1\eps_2)}^k} \int
 \frac{\Delta(0)\Delta(\eps_1 + \eps_2)}{\Delta(\eps_1)\Delta(\eps_2)} \prod_{j = 1}^k \frac{\dd \phi_{j}}{2\pi i} \\
&\times \prod_{j=1}^k \frac{\phi_{j}^2 \left(\phi_{j}^2 - ({\eps/2)}^2 \right)}{\Pol(\phi_{j} + \eps/2) \Pol(\phi_{j} - \eps/2)}.
\end{aligned}
\end{equation}
The factor $2^k k!$ is the order of the Weyl group of $Sp(k)$.
The instanton factor $\Lambda_5$ should be scaled as:
$$
{\Lambda}_5 \propto ({\b}{\Lambda}_4)^{2(N-2)} = ({\b}{\Lambda}_4)^{2h^{\vee}_{SO(N)}}
$$
Note that for the $B_n$ case the last line in \Ref{ZSO}
can be written as follows:
$$
\prod_{j=1}^k \frac{\phi_{j}^2}{\tilde{\Pol}(\phi_{j} + \eps/2)\tilde{\Pol}(\phi_{j} - \eps/2)}
$$
where $\displaystyle\tilde{\Pol}(x) = \prod_{l=1}^N (x^2 - a_l^2)$.

\subsection{The contour integrals}

The formulae \Ref{ZSO},\Ref{Zsp}
for the instanton partition function which we have got are
the multiple contour integral of a ratio of
two terms. They can be viewed as  the result of
integrating out the bosonic and fermionic auxiliary fields in
the finite dimensional representation of the functional integral
(see  \cite{GeneralizedInst,SWfromInst}).

The partition function can be expanded in $\Lambda \equiv \Lambda_4$ as follows
$$
Z(a,{\Lambda}_4 ;\eps_1,\eps_2) = Z_{\mr{pert}}(a,{\Lambda}_4 ;\eps_1,\eps_2)
\sum_{k=0}^\infty \Lambda^{2 h^\vee k} Z_k(a;\eps_1,\eps_2).$$
Each $Z_k(a;\eps_1,\eps_2)$ can be calculated by evaluating the integrals
by residues. These residues correspond to
the quadruples $B_1,B_2,I$ and $J$
which are invariant under the action of the
(maximal torus of the) gauge group, the dual group and the Lorentz group.
The condition of stability gives us the denominator.
The same procedure being applied to the
ADHM equations provides the numerator.
See  \cite{SWandRP} for more details. We shall not discuss the details of the
evaluation of these integrals here, as we shall only need a
saddle point in order to extract the prepotential. For future reference
we give here the analysis of the residues (we use the language
of fixed points). The reader interested in the prepotential {\em solo}
can skip the next subsection.


\subsubsection*{The poles}

\noindent\underline{Algebra $C_N$}.
The condition to be satisfied is:
\begin{equation}
\label{SpCondition}
\begin{aligned}
\left[B_1,\phi\right]&= i\eps_1 B_1, &  \left[B_2,\phi\right] &= i\eps_2 B_2, \\
-\phi I + I a &= -\frac{i\eps}{2}I,\qquad & -aJ + J\phi &= - \frac{i\eps}{2}J
\end{aligned}
\end{equation}
to obtain the denominator and
$$
\begin{aligned}
\left[ \mu_\Compl,\phi\right] &= i\eps \mu_\Compl, \qquad & \left[ \mu_\Real,\phi\right] &= 0
\end{aligned}
$$
to obtain the numerator. Here $\eps = \eps_1 + \eps_2$.

For the matrices $K$ and $K'$ from \Ref{IJSpN} we get
$$
\begin{aligned}
-\tilde{a} K + K \phi &= - \frac{i\eps}{2} K, &\qquad \tilde{a}K' + K' \phi = -\frac{i\eps}{2} K'.
\end{aligned}
$$
In order to diagonalize $\phi$ let
us introduce the following $k \times k$ matrix
\begin{equation}
\label{MatrixU}
U = \diag \left\{
\frac{1}{\sqrt{2}}\left(
\begin{array}{cc}
1 & 1 \\
-i & i
\end{array}
\right),\dots,\frac{1}{\sqrt{2}}\left(
\begin{array}{cc}
1 & 1 \\
-i & i
\end{array}
\right),\boxtimes \right\}
\end{equation}
where $\boxtimes = 1$ for odd $k$ and is absent for even $k$. One sees that
$$\tilde{\phi} = U^\dag \phi U
= \diag\left\{i \phi_1, - i\phi_1, i\phi_2,-i\phi_2,\dots,i \phi_n,-i\phi_n,\diamondsuit \right\}$$
and for $\tilde{K} = K U$ and $\tilde{K}' = K' U$ we obtain
$$
\begin{aligned}
-\tilde{a} \tilde{K} + \tilde{K} \tilde{\phi} &= - \frac{i\eps}{2} \tilde{K}, & \qquad
\tilde{a}\tilde{K}' + \tilde{K}' \tilde{\phi} = -\frac{i\eps}{2} \tilde{K}'.
\end{aligned}
$$
Hence the condition that the matrix elements of $K$ and $K'$ does not vanish for even $k$ are
$$
\begin{aligned}
- a_l + \phi_{i} + \frac{\eps}{2} &= 0, \quad & -a_l -  \phi_{i} + \frac{\eps}{2} &= 0, \\
 a_l + \phi_{i} + \frac{\eps}{2} &= 0, \quad & a_l -  \phi_{i} + \frac{\eps}{2} &= 0
\end{aligned}
$$
where $l = 1,\dots,n$ and $i = 1,\dots,N$.
For odd $k$ we have a supplementary pair of conditions
$$
\begin{aligned}
-a_l + \frac{\eps}{2} &= 0, \qquad & a_l + \frac{\eps}{2} &= 0.
\end{aligned}
$$
Now let us obtain condition under which the nontrivial solution for $B_s$, $s = 1,2$ are possible. Consider the case of even $k$. After examinating the first line of \Ref{SpCondition}
 we arrive to the following equations for $i \neq j$:
$$
\begin{aligned}
-\eps_s - \phi_{i} -  \phi_{j} &= 0, \qquad & -\eps_s + \phi_{i} +  \phi_{j} &= 0, \\
-\eps_s - \phi_{i} +  \phi_{j} &= 0, \qquad & -\eps_s + \phi_{i} -  \phi_{j} &= 0. \\
\end{aligned}
$$
And for $i = j$
$$
\begin{aligned}
-\eps_s &= 0, & -\eps_s - 2\phi_{i} &= 0, & \ \ &\mbox{and}\ \ & -\eps_s + 2\phi_{i} = 0.
\end{aligned}
$$
For odd $k$ we get additional equations:
$$
\begin{aligned}
-\eps_s &= 0, & -\eps_s + \phi_{i} &=0, & \ \ &\mbox{and} \ \ &\eps_s + \phi_{i} &=0.
\end{aligned}
$$
For $\mu_\Compl$ and $\mu_\Real$ we get the same equations with the replacement
$\eps_s \mapsto \eps$ and $0$ respectively.


\noindent\underline{Algebras $B_n$ and  $D_n$}
Now let us consider the group $SO(N)$.
Now we have to rewrite the equations \Ref{SpCondition} in terms of the building blocks for matrices $B_s$ \Ref{SOBB}, $M_\Compl$ and $N_\Compl$ \Ref{SOMN}. The stability conditions are ($s = 1,2$):
$$
\begin{aligned}
P_s \tilde{\phi} - \tilde{\phi} P_s &= i\eps_s P_s, & M_\Compl \tilde{\phi} - \tilde{\phi} M_\Compl = i\eps M_\Compl, \\
Q_s \tilde{\phi} + \tilde{\phi} Q_s &= i\eps_s Q_s, & N_\Compl \tilde{\phi} + \tilde{\phi} N_\Compl = i\eps N_\Compl, \\
-Q_s' \tilde{\phi} - \tilde{\phi} Q_s' &= i\eps_s Q_s', &  -N_\Compl' \tilde{\phi} - \tilde{\phi} N_\Compl' = i\eps N_\Compl'.
\end{aligned}
$$
After introducing the $N \times N$ dimensional version of \Ref{MatrixU} we get the following conditions for $\tilde{K} = U^\dag K$ and $\tilde{K}' = U^\dag K'$,
where $K$ and $K'$ are defined in \Ref{SOYZIJ}:
$$
\begin{aligned}
-\tilde{a}\tilde{K} + \tilde{K}\tilde{\phi} &= -\frac{i\eps}{2}\tilde{K}, & \qquad
-\tilde{a}\tilde{K}' - \tilde{K}'
\tilde{\phi} &= -\frac{i\eps}{2}\tilde{K}',
\end{aligned}
$$
where $\tilde{a} = U^\dag a U = \diag\{ i a_1, - ia_1,ia_2,-ia_2,\dots,\boxtimes\}$, $\boxtimes = 0$ for odd $N$ and is absent for even $N$.

For even $N$ the   condition to find nontrivial $\tilde{K}$ and $\tilde{K}'$ is:
$$
\begin{aligned}
-a_l + \phi_{i} &= -\frac{\eps}{2}, &\qquad +a_l + \phi_{i} &= - \frac{\eps}{2}, \\
-a_l - \phi_{i} &= -\frac{\eps}{2}, & \qquad +a_l - \phi_{i} &= - \frac{\eps}{2},
\end{aligned}
$$
where $l = 1,\dots,n$ and $i = 1,\dots,k$. For odd $N$ we have as well
$$
\begin{aligned}
\phi_{i} &= -\frac{\eps}{2},\qquad & -\phi_{i} &= -\frac{\eps}{2}.
\end{aligned}
$$
The conditions to find nontrivial $P_s, Q_s, Q_s'$ for all $N$ are:
$$
\begin{aligned}
\phi_{i} - \phi_{j} & = \eps_s, & \qquad i,j &= 1,\dots,k, \\
\phi_{i} + \phi_{j} &= \eps_s, & \qquad i &< j, \\
-\phi_{i} - \phi_{j} &= \eps_s, & \qquad i &< j.
\end{aligned}
$$
The same procedure applied to $\mu_\Compl$ gives:
$$
\begin{aligned}
\phi_{i} - \phi_{j} &= \eps, & \qquad i,j &= 1,\dots,k, \\
\phi_{i} + \phi_{j} &= \eps, & \qquad i &\leq j, \\
-\phi_{i} - \phi_{j} &= \eps, & \qquad i &\leq j.
\end{aligned}
$$


\section{Seiberg-Witten data}
\label{SW}

In this section we apply the method developed in  \cite{SWandRP} to obtain the Seiberg-Witten data for the prepotential.

It is known from  \cite{SWfromInst} that in $\eps_1,\eps_2 \to 0$ limit the leading Laurent expansion of $\ln Z(a,\Lambda;\eps_1,\eps_2)$ gives the prepotential. The main contribution comes form the instanton number which
has the same order as $\dfrac{1}{\eps_1\eps_2}$.
Hence we have to consider also $k\to\infty$ limit,
taken in such a way that $\eps_1\eps_2 k$ stays finite.

\subsubsection*{A trivial model example}

To illustrate the phenomenon, where the series is evaluated by the saddle
point we take the following trivial example:
\begin{equation}
Z ( {\Lambda}, {\hbar} ) = \sum_{k=0}^{\infty}
\frac{1}{k!} \left(\frac{\Lambda}{\hbar}\right)^k
\label{ex}
\end{equation}
Suppose ${\Lambda}/{\hbar} \in {\bR}_{+}$ and ${\Lambda} \gg \hbar$. Then
the series is dominated by the single term, where $k  = k_* \sim
{\Lambda}/{\hbar}$. Stirling's formula gives:
$$
Z( {\Lambda}, {\hbar} )  \sim \ex^{k_{*}} \sim {\exp}\frac{\Lambda}{\hbar}
$$
Now this formula can be analytically continued to aritrary $\Lambda \in
{\bC}$, and by expanding the answer in powers of $\Lambda$ we get correctly
the terms in the original series for small $k$.


\subsubsection*{$SO(N)$ case}

To begin with we note that using the formula
$$
\frac{\Delta(0)\Delta(\eps_1 + \eps_2)}{\Delta(\eps_1)\Delta(\eps_2)} = \exp \left\{ {\left.\eps_1\eps_2\frac{\dd^2}{\dd t^2} \ln \Delta(t)\right|}_{t = 0} + O(\eps^2) \right\}
$$
we can rewrite the expression for $Z_k(a;\eps_1,\eps_2)$ \Ref{ZSO} as follows
$$
\begin{aligned}
Z_k(a;\eps_1,\eps_2) &\sim \exp\left\{ -2 \eps_1\eps_2\sum_{i<j}
\left( \frac{1}{{(\phi_{i} - \phi_{j})}^2}  + \frac{1}{{(\phi_{i} + \phi_{j})}^2} \right) \right. \\
&+ \left. 4 \sum_{i = 1}^k \ln |\phi_{i}|  - 2 \sum_{i = 1}^k \sum_{l=1}^n \ln |\phi_{i}^2 - a_l^2| - 2\chi \sum_{i=1}^k
\ln|\phi_{i}| \right\}.
\end{aligned}
$$

Let us now introduce the density of $\phi_{i}$'s:
\begin{equation}
\label{rhoSO}
\rho(x) = \eps_1\eps_2 \sum_{I=1}^k \Big( \delta(x - \phi_{i}) + \delta(x + \phi_{i})\Big).
\end{equation}
We see immediately that the normalization of the density function is such that the integral $\int_\Real \rho(x)\dd x$ is finite in the limit in hand. It worth noting that the density function is even: $\rho(x) = \rho(-x)$.

Now we can rewrite \Ref{ZSO} as follows:
$$
Z_k(a;\eps_1\eps_2) \sim \exp \left\{ -\frac{1}{\eps_1\eps_2} \mathcal{E}_1 \right\}
$$
where the ``free energy'' $\mathcal{E}_\Lambda$ is
$$\mathcal{E}_\Lambda =  \frac{1}{2}\vpint \dd x \dd y \frac{\rho(x)\rho(y)}{{(x-y)}^2} + \int \dd x \rho(x) \ln\left|\frac{\Pol(x)}{\Lambda^{2n + \chi}}\right| -  2 \int \dd x \rho(x) \ln\left|\frac{x}{\Lambda}\right|.
$$

It worth noting the following identity
\begin{equation}
\label{ZviaE}
\Lambda^{2kh^\vee} Z_k (a;\eps_1,\eps_2) \sim \exp\left\{ -\frac{1}{\eps_1\eps_2}\mathcal{E}_\Lambda \right\}.
\end{equation}

To go further let us introduce the profile function (see  \cite{SWandRP} for the explanation of such a name) defined for both even $N$ and odd $N$ cases  as follows:
\begin{equation}
\label{defProf}
f(x) = -2\rho(x) + \sum_{l=1}^n \Big(|x- a_l| + |x+a_l| \Big),
\end{equation}
and the kernel:
$$
\k_\Lambda(x) = \frac{1}{2} x^2 \left( \ln \left|\frac{x}{\Lambda}\right| -\frac{3}{2} \right).
$$
The kernel function is defined at
the $0$
 by continuity, i.e. $\k_\Lambda (0) = 0$.

Then the free energy can be rewritten as follows:
\begin{equation}
\label{FreeEn}
\begin{aligned}
\mathcal{E}_\Lambda &= - \frac{1}{8}\int  \dd x \dd yf''(x)f''(y) \k_\Lambda(x-y) + \left( 1 - \frac{\chi}{2}\right) \int \dd x f''(x)\k_\Lambda(x) \\
&+ \left\{ \sum_{l < m} \Big( 2 \k_\Lambda(a_l - a_m) + 2\k_\Lambda(a_l + a_m) \Big) + \chi \sum_{l=1}^n 2\k_\Lambda(a_l) \right\}.
\end{aligned}
\end{equation}

The term in the curly brackets can be recognized as the expression for the perturbative (1-loop) prepotential \Ref{pertPrep}  for both $SO(2n)$ and $SO(2n + 1)$.

We can express the full (quantum part of the) prepotential as follows:
\begin{multline}
\label{ActSO}
\Prep_{\mr{quant}}(a,\Lambda) = \Prep_{\mr{pert}}(a,\Lambda) + \Prep_{\mr{inst}}(a,\Lambda) \\
= -\frac{1}{8} \int \dd x \dd y f''(x)f''(y)\k_\Lambda(x-y) + \left( 1 - \frac{\chi}{2}\right) \int \dd x f''(x)\k_\Lambda(x).
\end{multline}

The main contribution to this expression is given by the saddle point of this functional  \cite{SWandRP}. The support of the maximizer is the union of the intervals $[\alpha^-_{l,\pm},\alpha^+_{l,\pm}] \ni \pm a_l$. The variation should be taken with respect to the functions satisfying the following condition:
\begin{equation}
\label{fCond}
\frac{1}{2}\int_{\alpha^-_{l,\pm}}^{\alpha^+_{l,\pm}}  x f''(x) \dd x= \pm a_l.
\end{equation}

Instead of solving this variational problem we note that the expression \Ref{ActSO} for the prepotential is (the factor $1/2$ of) the expression considered in  \cite{SWandRP} which corresponds to the case of the gauge group $SU(2N)$ with $4 - 2\chi$ massless matter multiplets and with the vacuum expectation of the Higgs field $\langle \phi \rangle = \diag\{ i a_1, - ia_1,\dots,ia_n,-ia_n\}$.

Already this observation enables us, in particular, to rederive the theorem, stated in  \cite{SOandSp,LecturesSWandIntegr}, which says that the prepotential for the pure Yang-Mills theory for the gauge group $SO(2n)$ equals to the one half of the prepotential for $SU(2n)$ coupled to the 4 massless matter multiplets and that the prepotential for the pure Yang-Mills theory for $SO(2n + 1)$ equals one half of the prepotential for $SU(2n)$ coupled to the 2 massless matter multiplets.

To write down the solution let us introduce the following function (the primitive of the resolvent of the maximizer)
\begin{equation}
\label{Fdef}
F(z) = \frac{1}{4\pi i} \int_\Real f''(x)\ln \left(\frac{z - x}{\Lambda}\right)\dd x.
\end{equation}
After the conventional reparametrization $F(z) = \ln w(z)$ the maximal value of the righthand side of \Ref{ActSO} can be expressed as follows. We consider the algebraic curve
\begin{equation}
\label{SOCurve}
w + \frac{\Lambda^{4n + 2\chi - 4} z^{4 - 2\chi}}{w} = \prod_{l=1}^n (z^2 - \alpha_l^2)
\end{equation}
where $\pm i\alpha_l$ are the classical eigenvalues of $\phi$. The quantum eigenvalues $\pm i a_l$ can be expressed as follows (we use \Ref{Fdef} and \Ref{fCond}):
$$
\pm a_l = \frac{1}{2\pi i} \int_{A_{l,\pm}} z \dd F = \frac{1}{2\pi i} \int_{A_{l,\pm}} z \frac{\dd w}{w}
$$
whereas the prepotential can be expressed as
\begin{equation}
\label{factor2}
2 \times \frac{\pd \Prep_{\mr{quant}}(a,\Lambda)}{\pd a_l} = \int_{B_l} z \dd F = \int_{B_l} z \frac{\dd w}{w}.
\end{equation}
The description of cycles $A_{l,\pm}$ are $B_l$ the following. Let $\alpha_{l,\pm}^+$ and $\alpha_{l,\pm}^-$ are solutions of the equation $\displaystyle \prod_{l=1}^n (z^2 - \alpha_l^2) = 2 \Lambda^{2n}$ such that $\pm\alpha_l \in [\alpha^-_{l,\pm},\alpha^+_{l,\pm}]$. When $A_{l,\pm}$ goes around the cut $[\alpha_{l,\pm}^-,\alpha_{l,\pm}^+]$. As for $B_l$, we can take a contour which goes from $\alpha_{l,+}^+$ to $\alpha_{l+1,+}^-$ on the on leaf of \Ref{SOCurve} and closes on the another.

Comparing to \Ref{SWdata} we see that the curve $\Ref{SOCurve}$ and the differential $\displaystyle \lambda = z\dd F = z\frac{\dd w}{w}$ are essentially the Seiberg-Witten data we are looking for.

The following remark should be made. The shift of the density function \Ref{defProf} is not uniquely defined.
We could, for example,
add to the profile function the term $\zeta |x|$ with some
positive integer $\zeta$. The price we pay is the shift $\chi \mapsto \chi - \zeta$. It follows that the curve \Ref{SOCurve} becomes
$$
w + \frac{\Lambda^{4n + 2\chi - 4} z^{4- 2\chi + 2\zeta}}{w} = z^\zeta \prod_{l=1}^n(z^2 - \alpha_l^2).
$$
Since physically nothing is changed the new curve should be equivalent to the old one. This is, indeed, the case and the following transformation $w \mapsto w z^\zeta$ does
the job of relating the curves.
We see that this transformation does not change the cohomology class of $ \lambda$.

Both the curve and the differential is in the agreement with other results.


\subsubsection*{$Sp(N)$ case}

Let us repeat the same steps to derive the expression for the Seiberg-Witten curve and for the differential in the case of the group $Sp(N)$.

The expression \Ref{Zsp} for $Z_k(a;\eps_1,\eps_2)$ can be rewritten as follows:
$$
\begin{aligned}
Z_k(a;\eps_1,\eps_2) &\sim \exp \left\{ -2 \eps_1 \eps_2 \sum_{i <j}
\left( \frac{1}{{(\phi_{i} - \phi_{j})}^2} +  \frac{1}{{(\phi_{i} + \phi_{j})}^2}\right) \right. \\
&+ \left. 2\chi\eps_1\eps_2 \sum_{i=1}^n \frac{1}{\phi_{i}^2} - 4 \sum_{i=1}^n \ln|\phi_{i}| -
2\sum_{i=1}^n \sum_{l=1}^N \ln|\phi_{i}^2 - a_l^2|
 \right\}
\end{aligned}
$$
We recall that when we consider $Sp(N)$ gauge group we denote $k = 2n + \chi$, $n = [k/2]$.

After having introduced the density
$$
\rho(x) = \eps_1\eps_2 \sum_{i=1}^n \Big( \delta(x - \phi_{i}) + \delta(x + \phi_{i}) \Big)
$$
we can rewrite $Z_k(a;\eps_1,\eps_2)$ as \Ref{ZviaE} where
$$
\mathcal{E}_\Lambda = \frac{1}{2} \vpint \dd x\dd y \frac{\rho(x)\rho(y)}{{(x - y)}^2} + \int \dd x \rho(x) \ln \left| \frac{\Pol(x)}{\Lambda^{2n}}\right| + 2 \int \dd x \rho(x) \ln \left|\frac{x}{\Lambda}\right|.
$$
It worth noting that the sum of the inverse squares of $\phi_{i}$ gives the subleading contribution and therefore is dropped.

The definition of the profile function is
\begin{equation}
\label{profSp}
f(x) = - 2 \rho(x) + \sum_{l=1}^N \Big( |x-a_l| + |x + a_l|\Big) + 2 |x|.
\end{equation}

The analog of the expression \Ref{ActSO} is
\begin{multline}
\label{ActSp}
2\pi i \Prep_{\mr{quant}} = 2\pi i \Prep_{\mr{pert}}(a,\Lambda) + 2\pi i \Prep_{\mr{inst}}(a,\Lambda) = \\
- \frac{1}{8}\int \dd x\dd y f''(x)f''(y)\k_\Lambda(x-y)
\end{multline}
where
$$
2\pi i\Prep_{\mr{pert}}(a,\Lambda) = \sum_{l<m} 2\k_\Lambda(a_l - a_m) + \sum_{l<m} 2\k_\Lambda(a_l + a_m) + \sum_{l=1}^N 2\k_\Lambda(2a_l)
$$
in agreement with \Ref{pertPrep}.

We see that the problem of finding the prepotential for the group $Sp(N)$ is equivalent to the problem of finding the prepotential for $SU(2N + 2)$ where the vacuum expectation of the Higgs field is
\begin{equation}
\label{SpHiggs}
\langle \phi \rangle = \diag\{ ia_1,-ia_1,\dots,ia_N,-ia_N,0,0\}.
\end{equation}
In fact, in  \cite{SOandSp,LecturesSWandIntegr} this statement was formulated differently. However we point out that one can get different forms of this statement by taking into account the remark following after \Ref{SOCurve}. Roughly speaking, each zero eigenvalue of $\phi$ ``worth'' two massless multiplets. Hence if we add 4 massless multiplets to the theory, the prepotential becomes equal to the prepotential for $SU(2N)$.

In fact we can do more, we can write down the expression for the curve using the results of  \cite{SWandRP}. For $F(z)$ and $w(z)$ defined in \Ref{Fdef} we have (see the appendix \ref{Spgeom} for the details)
$$
w + \frac{\Lambda^{4N + 4}}{w} = z^2 \prod_{l=1}^N (z^2 - \alpha_l^2) + 2\Lambda^{2N + 2}.
$$
The differential will be defined as in the $SU(2N+2)$ (and also $SO(N)$) case as $\displaystyle\lambda = z \dd F = z\frac{\dd w}{w}$.


\section{Instanton corrections}
\label{check}

In this section we compute some lower $k$ instanton correction in order to perform a consistency check and to compare our results against the
existing in the literature.


\subsection{Consistency check}

Let us perform a check of our results.

The curves and the differentials for the $SO(N)$ and $Sp(N)$ theories are obtained in $k\to\infty$ limit.
However, they give the right answer even in the case of small $k$.
This is the essence of the Seiberg-Witten theory.
Hence, for example, the 1-instanton correction can be extracted form them.
But for small $k$ the the instanton corrections can be calculated straightforwardly using \Ref{ZSO} and \Ref{Zsp}. Thus we can compare former and later expressions.
Note that this is highly nontrivial check since, for example,
for $Sp(N)$ the 1-instanton correction comes from the off-integral
 term which does not participate at all in the derivation of \Ref{ActSp}.

Luckily for us, the hard part of the job, the extraction of
the 1-instanton corrections from the curves, is already done (see, for example,  \cite{SOandSp,EllipticMod}).
Let us cite the result for $\Prep_1(a)$ (see \Ref{instPrep}):
$$
\begin{aligned}
2 \pi i\Prep_1(a) &=- 2{(-1)}^N \prod_{l=1}^N \frac{1}{a_l^2} & &\mbox{for $C_N$,} \\
2 \pi i\Prep_1(a) &= \frac{1}{4}\sum_{l=1}^n \frac{1}{\prod_{k\neq l} {(a_k^2 - a_l^2)}^2} & &\mbox{for $B_n$,} \\
2\pi i\Prep_1(a) &=  \frac{1}{4}\sum_{l=1}^n\frac{a_l^2}{\prod_{k\neq 1} {(a_k^2 - a_l^2)}^2} & &\mbox{for $D_n$.}
\end{aligned}
$$
To compute the integrals in \Ref{ZSO} and \Ref{Zsp} we need the prescription to go around the poles.
It can be obtained by careful analysis of the integration out of the bosonic auxiliary fields
 \cite{DeformationInstanton,GeneralizedInst}
and consists of the shift
$\eps_s \mapsto \eps_s + i 0$. As we argued below \Ref{projrat}, this
prescription can be understood very simply from the five dimensional
perspective, as the condition of the convergence of the character.

Note that we don't need any closure prescription.
Indeed, whether we close the contour in
the upper or in the lower half-plane the result will be the same since the residue of infinity vanishes.

Taking this remark into account we see that our results are in the agreement with the formulae cited above (up to an overall factor which can be absorbed to the definition of $\Lambda$).


\subsection{$Sp(N)$ instanton corrections}

Looking at \Ref{Zsp} we see that in the case of $Sp(N)$ to obtain $k$th instanton correction we have to compute only $[k/2]$tiple integral. In particular to get $\Prep_1(a)$, $\Prep_2(a)$ and $\Prep_3(a)$ we can compute a single $\dd \phi$ integral.

For $Z_1(a;\eps_1,\eps_2)$, $Z_2(a;\eps_1,\eps_2)$ and $Z_3(a;\eps_1,\eps_2)$ we have:
$$
\begin{aligned}
Z_1(a;\eps_1,\eps_2) &= - \frac{1}{2\eps_1\eps_2} \prod_{l=1}^N \frac{1}{a_l^2- {(\eps / 2)}^2}, \\
Z_2(a;\eps_1,\eps_2) &= -\frac{1}{2\eps_1\eps_2} \sum_{l=1}^N \Big(S_l(a_l^-)D(a_l^-) + S_l(a_l^+)D(a_l^+) \Big) \\
&+ \frac{1}{8{(\eps_1\eps_2)}^2} \frac{\eps_2 T(\eps_1/2) - \eps_1 T(\eps_2/2)}{\eps_2 - \eps_1}, \\
Z_3(a;\eps_1,\eps_2) &=\frac{1}{4{(\eps_1\eps_2)}^2} \prod_{l=1}^N \frac{1}{a_l^2- {(\eps / 2)}^2} \\
&\times\left\{ \sum_{l=1}^N \Big( S_l(a_l^-)D(a_l^-)E(a_l^-) +  S_l(a_l^+)D(a_l^+)E(a_l^+)\Big)\right. \\
&+\frac{1}{12\eps_1\eps_2}\left[ -\frac{T(\eps_1/2) + T(\eps_2/2)+2T(\eps_1)+2T(\eps_2)}{6} \right.\\
&+(\eps_1 + \eps_2)\left(-\frac{2\left(T(\eps_1)-T(\eps_2)\right)+5\left(T(\eps_1/2)-T(\eps_2/2)\right)}{\eps_1-\eps_2}\right.\\
&\left.\left.\left. -\frac{8\eps_1\left(T(\eps_2/2)-T(\eps_1)\right)}{3(\eps_1-2\eps_2)}-\frac{8\eps_2\left(T(\eps_1/2)-T(\eps_2)\right)}{3(\eps_2-2\eps_1)}\right)\right]\right\}
\end{aligned}
$$
where
$$
\begin{aligned}
a_l^\pm &= a_l \pm \frac{\eps_1+\eps_2}{2}, \\
D(t) &=\frac{1}{\left[4t^2 - \eps_1^2\right]\left[4t^2 - \eps_2^2\right]}, \\
E(t) &= \frac{t^2(t^2-(\eps_1+\eps_2)^2)}{(t^2-\eps_1^2)(t^2-\eps_2^2)},\\
S_l(t) &= \frac{1}{4a_lt} \prod_{l\neq k} \frac{1}{\left[{(t+\eps/2)}^2 - a_k^2\right]\left[{(t-\eps/2)}^2- a_k^2\right]},\\
T(t) &= \frac{1}{\Pol(\eps/2 + t)\Pol(\eps/2 - t)}.
\end{aligned}
$$
Using the definition we get the prepotential:
$$
\begin{aligned}
2\pi i\Prep_1(a) &= -\frac{1}{2}\prod_{l=1}^N \frac{1}{a_l^2}, \\
2\pi i\Prep_2(a) &= - \frac{1}{16}\left(\sum_{l=1}^N \frac{\tilde{S}_l(a_l)}{a_l^4} + \frac{1}{4}{\left.\frac{\pd^2 \tilde{T}(t)}{\pd t^2}\right|}_{t=0}\right), \\
2\pi i\Prep_3(a) &= -\frac{1}{16}\prod_{l=1}^N\frac{1}{a_l^2} \left( \sum_{l=1}^N \frac{\tilde{S}(a_l)}{a_l^6} + \frac{1}{144}{\left.\frac{\pd^4\tilde{T}(t)}{\pd t^4}\right|}_{t=0}\right)
\end{aligned}
$$
where the tilde over $S$ and $T$ means that we set $\eps=0$ in the definition of these functions.

For the case $N=1$, that is, for the group $Sp(1) = SU(2)$ we have the following expression for the prepotential:
$$
2\pi i\Prep_{\mr{inst}}(a,\Lambda) =  -\frac{\Lambda^4}{2a^2} - \frac{5\Lambda^8}{64a^6} -\frac{3\Lambda^{12}}{64a^{10}}+ O(\Lambda^{16})
$$
which coincides with both Seiberg-Witten  \cite{HKP-SU} and direct computations  \cite{SWfromInst} for $SU(2)$. $Z_k(a;\eps_1,\eps_2)$, $k=1,2,3$ can also be checked against the corresponding quantities for $SU(2)$  \cite{SWfromInst}.

\section{Conclusions}

In this paper we have derived the prepotentials of the low-energy
effective theory for ${\CN}=2$ supersymmetric Yang-Mills theories
with gauge groups $SO(N)$ and $Sp(N)$.
We have done so by expressing partition function of
the four dimensional gauge theory in $\Omega$-bacgkround
as an infinite sum (over instanton numbers) of contour integrals.
For approprate values of the vacua parameters this sum is saturated
by the contribution of a single (large) instanton number, and the
corresponding integral is saturated by the contribution of the
single saddle point. This saddle point corresponds to some sort
of eigen-value distribution, which is encoded in an algebraic
curve, Seiberg-Witten curve.

Our approach {\em does not} require the resolution of singularities
of the instanton moduli space. However, at present, it requires
the knowledge of the ADHM construction. This is an obstacle which does not
allow, at the moment, to extend our methods to the case of $E,F,G$ gauge
groups.

We also note that we did not discuss theories with matter in this paper.
As far as the asymptotically free theories are concerned, the matter
is easily treated within our approach. The results of  \cite{SWandRP}
give hope that the asymptotically conformal theories should also be
possible to include in this framework.

\underline{Note \ added.} When this paper was at the stage of preparation
we received the preprint  \cite{Marcos} which discusses the evaluation of
the integrals \Ref{Zsp}, \Ref{ZSO}, by  adopting the ${\ve} + i0$ prescription
of  \cite{DeformationInstanton}. We point out that our five dimensional
definition gives an unambiguous definition of all of these integrals,
including
 \Ref{ZSO} which apparently are difficult to treat by the methods of
 \cite{Marcos}. We also point out that the conjecture of  \cite{SWandRP}
which expresses the {\em dual} instanton partitions as matrix
elements of the affine Lie algebras gives a hint about the residues
of the integrals \Ref{ZSO},\Ref{Zsp} and their up-to-date unknown $E_{6},E_{7}, E_{8}, F_{4} ,G_{2}$
analogues.

\section*{Acknowledgements}

We are grateful to  A.~Khoroshkin, M.~Kontsevich, A.~Losev, and
especially A.~Okounkov  for pleasant discussions. NN is
also grateful to H.~Nakajima for giving an inspiring seminar in Paris.
NN is
grateful to B.~Julia for organizing such a seminar.
The research of NN is partly supported by {\cyr RFFI} grant 03-02-17554
and
by the grant {\cyr NSh}-1999.2003.2 for scientific schools.


\appendix

\section{Seiberg-Witten geometry for $Sp(N)$ case}
\label{Spgeom}

In this appendix we derive in some details the Seiberg-Witten geometry for the $Sp(N)$ case. In section \ref{SW} we have pointed out that the variational problem for $Sp(N)$ formally coincides with the problem for $SU(2N+2)$ with two zero Higgs field vevs. However, the method of  \cite{SWandRP} should be slightly modified\footnote{it was kindly pointed out to us by N. Wyllard and M. Mari\~no}. The reason is that in the case at hand one of the crucial conditions for the solution of variational problem to be valid, $|a_l - a_m| \gg \Lambda$ for $l\neq m$, is spoiled. Here we describe the solution for the case where the Higgs vacuum expectations are given by \Ref{SpHiggs}.

\begin{figure}[h]
\includegraphics[width=\textwidth]{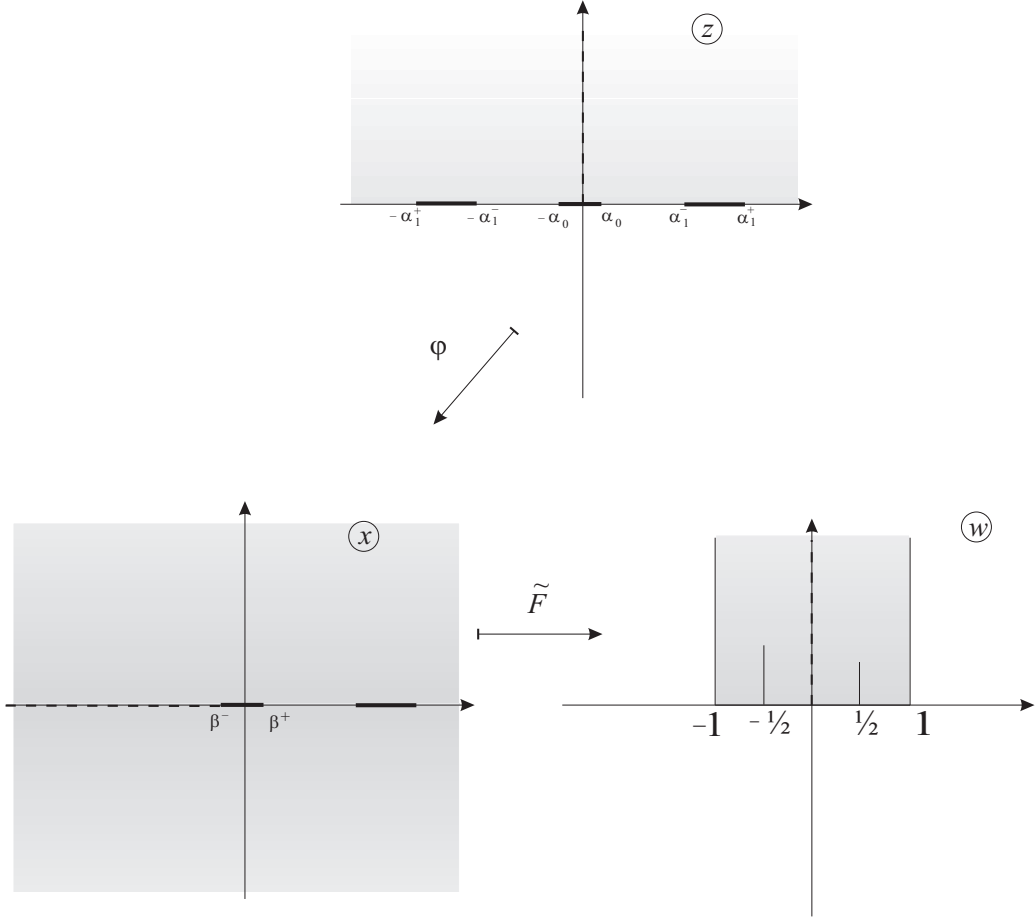}
\caption{Conformal map for $Sp(N)$, $N=1$}\label{SpPicture}
\end{figure}

Let $[-\alpha_0,\alpha_0]$ be the part of support of $f(x)$ which contains zero. Then it follows directly form \Ref{profSp} that
$$\int_{-\alpha_0}^{\alpha_0} f'(x) \dd x = 4$$
(we recall that the righthand side equals 2 when all vevs are different). It follows that we should map the upper half plane to the domain traced at the $w$ plane of figure \ref{SpPicture}.

The map $\tilde{F}$ is given by\cite{SWandRP}
$$\tilde{F} = \frac{1}{2\pi} \arccos \frac{x \prod_{l=1}^N (x - \beta_l)}{2\Lambda^{N+1}}.$$
It maps the upper halfplane $x$ to the half of the domain at the plane $w$. According to the reflection principle it maps the whole complex plane to the whole domain. The endpoints of the interval $[\beta^-,\beta^+]$ satisfy the equation
$$\beta^\pm \prod_{l=1}^N (\beta^\pm - \beta_l) = \pm{(-1)}^N 2\Lambda^{N+1}.$$

The map $\varphi$ is given by the formula $z \mapsto  z^2/\Lambda + \beta^-$. Hence the map form $z$ plane to $w$ plane is given by
$$F = \tilde{F}\circ\varphi = \frac{1}{2\pi} \arccos \frac{\Pol(z)}{2\Lambda^{2N+2}}$$
where
$$
\begin{aligned}
\Pol(z) &= (z^2 + \Lambda \beta^-)\prod_{l=1}^N (z^2 + \Lambda \beta^- - \Lambda \beta_l) \\
& = z^{2N+2} + u_1 z^{2N} + \dots + u_N z^2 + \Lambda^{N+1}\beta^- \prod_{l=1}^N (\beta^-\ - \beta_l) \\
& = z^2 \prod_{l=1}^N (z^2 - \alpha_l^2) + 2{(-1)}^{N+1}\Lambda^{2N + 2}.
\end{aligned}
$$

It follows that after the redefinition $\Lambda \mapsto i\Lambda$ the Seiberg-Witten curve can be written as:
$$
w + \frac{\Lambda^{4N + 4}}{w} = z^2 \prod_{l=1}^N (z^2 - \alpha_l^2) + 2\Lambda^{2N + 2}.
$$



\providecommand{\bysame}{\leavevmode\hbox to3em{\hrulefill}\thinspace}
\providecommand{\MR}{\relax\ifhmode\unskip\space\fi MR }
\providecommand{\MRhref}[2]{%
  \href{http://www.ams.org/mathscinet-getitem?mr=#1}{#2}
}
\providecommand{\href}[2]{#2}

\end{document}